\documentclass{article}
\usepackage[preprint]{neurips_2025}
\usepackage{amsmath,amssymb}
\usepackage{enumitem}
\usepackage{booktabs}
\usepackage{graphicx}
\usepackage{subcaption}
\usepackage{algorithm}
\usepackage{algpseudocode}
\usepackage{placeins}
\usepackage{xurl}
\usepackage[colorlinks=true,linkcolor=blue,citecolor=blue,urlcolor=blue]{hyperref}

\title{Simulating American Option Prices with Dynamic Implied Volatility}

\author{
  Julia Sun,\; Zheyu Jin,\; Jiawei Zhang,\; and\, Jeffrey D.\ Varner\thanks{Corresponding author.} \\
  Robert Frederick Smith School of Chemical and Biomolecular Engineering \\
  Cornell University, Ithaca, NY 14853 \\
  {\fontencoding{T1}\selectfont\texttt{\{hhs68, zj276, jz2483, jdv27\}@cornell.edu}}
}

\begin{document}
\raggedbottom

\maketitle

\begin{abstract}
Valuing American options along simulated stock paths requires an implied volatility (IV) for every option on every date. A stock-return model alone does not provide it. We built a simulator that assigned IV to American options on simulated dates for any chosen stock model. We fitted parametric and neural IV surfaces to vendor option quotes for 31 tickers. Each surface predicted IV from moneyness and time to expiration. Stock paths came from a jump hidden Markov model with capped daily returns. Each day, every option's implied variance moved partway toward its surface prediction, following the form of the Heston variance equation. Random shocks tended to raise IV when the stock fell. A binomial tree converted IV into option values and price sensitivities. Neural surfaces fitted by sector or ticker matched the quotes more closely than one parametric surface. Their errors still varied by date and carried into dollar prices. Repricing identical stock paths under five ways of updating IV changed option values before expiration and the worst simulated short-position losses at a fixed horizon. Payoffs at expiration did not change. In forecasts of later Goldman Sachs and Eli Lilly option prices, stochastic IV did little better than fixed IV. Rerunning the forecasts with the realized stock paths pointed to stock prediction as a major source of option-price error. The hidden Markov model and an adaptive-volatility model predicted stock prices about as accurately as assuming no change, and tuning found no gain from predicting direction. The adaptive-volatility model improved predicted price ranges for Eli Lilly but not Goldman Sachs. The simulator supported reproducible comparisons of IV and stock assumptions but did not forecast better than simpler alternatives. Complete option histories and prices consistent across strikes are needed before simulated prices can replace market data.

\end{abstract}

\section{Introduction}\label{sec:introduction}

Stress tests and backtests of trading strategies can run on simulated stock paths as well as on historical prices. For a stock portfolio, the simulated prices directly determine the value of its holdings. Valuing options before expiration requires an additional pricing model. A common approach uses the option's implied volatility (IV), the volatility at which a pricing model, such as the Black--Scholes model~\citep{black1973}, reproduces the option's market price~\citep{latane1976}. A simulated stock path comes with no option prices from which to infer IV. The simulated stock's own volatility does not specify the IV of every option, because market IVs differ across strikes and expiration dates. For a given put or call exercised at expiration, the payoff depends only on the terminal stock price. Option positions, however, can be adjusted or closed before expiration. To value them with an IV-based pricing model, a simulator needs two parts. The first is a stock-return model that generates paths with features of observed returns, such as heavy tails and volatility clustering~\citep{cont2001}. Regime-switching models with jumps and generative neural networks have been used to model these features~\citep{wiese2020,jumphmm2025}. The second is a rule that assigns every option an IV and updates it as the simulated stock moves and expiration approaches.

Several families of models, once fitted to quotes, can assign an IV or a price to options that were not quoted. Parametric smile formulas, such as those derived from the stochastic alpha beta rho (SABR) model and the stochastic volatility inspired (SVI) form, describe the smile on a given date with a few parameters~\citep{hagan2002,gatheral2004}. Models of IV-surface dynamics describe how the surface's level, skew, and curvature change over time through common random drivers that move IVs across the surface~\citep{cont2002}. More recent surface-dynamics methods prevent static arbitrage, or inconsistencies among option prices on the same date, in simulated surfaces~\citep{cont2023}. When used as full pricing models, local- and stochastic-volatility models generate the stock and its volatility together within one pricing model~\citep{dupire1994,derman1994,heston1993}. Neural networks trained to imitate such models speed up pricing and calibration while retaining the pricing assumptions of the model they approximate~\citep{horvath2021,bayer2019}. We focused on how the choice of IV rule changes option values when the stock-return model and its simulated paths stay fixed. To make this comparison, we needed an update whose response to stock returns and additional random changes could be removed separately. Simple choices include holding each option's IV fixed or evaluating a previously fitted surface at its current moneyness and remaining time to expiration. With the surface parameters held fixed, neither rule introduces additional randomness once the stock path is given. Observed IV surfaces, however, also shift in ways the stock price does not explain~\citep{cont2002}. We therefore sought an update that takes the stock path as given, lets IV respond to its returns, and adds random IV changes beyond those responses. Running this update and simpler rules on the same stock paths would isolate how the choice of IV rule changes option values.

In this study, we built a simulator of American options using stock paths from a jump hidden Markov model. We fitted parametric and neural IV surfaces to vendor quotes before simulation, then held their parameters fixed. Each option's implied variance, the square of IV, followed a daily update adapted from the Heston variance equation~\citep{heston1993}. The update combined a pull toward the fitted surface's variance prediction with random shocks partly tied to stock returns, so declines tended to raise IV. A binomial tree converted IV into option prices~\citep{cox1979,leisen1996}. We asked four questions. First, we measured how well the surfaces matched reported IV. Neural surfaces fitted by sector or stock had lower overall fitting errors than the global parametric and neural surfaces. They still missed shifts in IV across dates, and sector surfaces had larger errors on withheld dates. Earnings-date inputs reduced some of those errors. Second, we compared IV rules on identical stock paths for Goldman Sachs and Eli Lilly options. The rules changed option values before expiration and the worst simulated short-position losses after ten trading days. With early assignment ignored, expiration payoffs were identical for the same option and stock path. Third, we checked prices for arbitrage across strikes. Some violated the tested conditions, mostly when IV followed the surface prediction directly; the surfaces had not been constrained to prevent these violations. We did not test arbitrage across expiration dates. Fourth, we compared mean simulated option prices with later quote midpoints. Our update reduced average absolute errors only slightly relative to fixed IV. Substituting observed stock paths substantially reduced option-price errors, indicating that stock prediction was a major source of error. The stock generator and alternatives that adapted to recent volatility or attempted to predict return direction all had point-forecast errors close to an unchanged-price benchmark. These results support controlled comparisons of IV rules on fixed stock paths, but neither improved forecasting nor a guarantee of arbitrage-free prices.

\section{Methods}\label{sec:methods}

\subsection{Simulator overview and stock paths}

The simulator valued options along stock paths in two phases. We first fitted IV surfaces to observed quotes, predicting IV from moneyness (strike divided by stock price) and time to expiration. We then generated complete stock paths and advanced along each path one trading day at a time. A binomial tree used the current stock price, remaining time, and each option's IV to calculate its price and price sensitivities (Greeks). For the next trading day, we updated each option's implied variance, the square of IV. The update combined a pull toward a variance target from the fitted surface with random shocks partly tied to the stock return over that step. We evaluated the target before advancing the stock price and time to expiration. Starting IVs came from the fitted surfaces, except in the forecast test, which used observed IVs. No subsequent quotes entered a simulation. Option prices never fed back into the surfaces or stock paths (Algorithm~\ref{alg:forward_sim}). Illustrations used short put and call positions on Goldman Sachs (GS) and Eli Lilly (LLY). To isolate the effect of IV, we repriced the same options on identical stock paths under five IV rules, a comparison we call the IV ablation. Separate tests assessed forecasts of option and stock prices.

The stock paths came from the Jump Hidden Markov Model (JumpHMM)~\cite{jumphmm2025}. We used models from the study that introduced JumpHMM, which fitted 424 tickers separately on 2{,}767 trading days from 2014 to 2024. The original fit assigned historical excess growth rates, the annualized log returns minus a risk-free rate, to 50 ordered bins and used bin membership as the state labels for fitting. Each state drew this rate from its own location-scale Student-$t$ distribution, the state's emission distribution. The models used a time step of $1/252$ year and a risk-free rate of zero, so their growth rates were annualized log returns. During jump episodes, selection of states from the lowest or highest bins replaced ordinary state transitions for a Poisson-distributed number of days. The original study tuned the frequency and duration of these episodes. Every simulation here used one ticker at a time.

For price simulation, we bounded the Student-$t$ emissions because exponentiating unbounded Student-$t$ log returns gives an infinite expected stock price, leaving mean call values and expected short-call losses without finite population values~\cite{cassidy2009}. The simulated growth rate on day $t$ was given by:
\begin{equation}\label{eq:truncated_emission}
 G_t=\mu_{k_t}+\sigma_{k_t} Z_t,\qquad
 Z_t\sim t_{\nu_{k_t}}\mid |Z_t|\le b,\qquad b=10,
\end{equation}
where $k_t$ is the hidden state on day $t$, $\mu_k$, $\sigma_k$, and $\nu_k$ are its fitted location, scale, and degrees of freedom, and $Z_t$ is a Student-$t$ draw restricted to $|Z_t|\le b$. The cutoff $b$ is measured in multiples of the state's scale. We replaced draws outside the bound with draws from the restricted distribution; no stock path was discarded. This change applied only during simulation, preserved the fitted parameters and state transitions, and gave stock prices finite means and variances at every fixed horizon. Prices followed $S_{t+1}=S_t\exp[(G_t+c)\Delta t]$ with $\Delta t=1/252$. To replace the drift inherited from 2014--2024, we estimated the constant growth shift $c$ separately for each ticker. The shifted annualized log-growth rates $G_t+c$ averaged an assumed 10\% in the estimation sample. The illustrations estimated $c$ from their sampled growth rates before converting them to prices. The IV ablation and forecast test used a separate pilot simulation, then drew fresh paths for scoring, called evaluation paths, which never influenced $c$. We chose $b=10$ and a wider $b=20$ sensitivity case in advance as stress assumptions, not by forecast performance. The wider-cutoff reruns repeated the stock and option simulations with the same options, seeds, and dates. They recomputed $c$ and the mean and standard deviation used to center and scale stock log returns for the stock-related IV shocks (Supplementary Tables~\ref{tab:emission_bounds}, \ref{tab:tail_sensitivity}, and~\ref{tab:tail_forecasts}).

\subsection{Option data and IV surfaces}

We fitted surfaces to the IVs reported in Alpaca Markets' indicative option feed, available with a free account. The feed derives from consolidated Options Price Reporting Authority (OPRA) data, but its quotes, midpoints, spreads, and IVs may differ from exchange quotes and executable prices~\cite{alpaca_option_feed}. Alpaca computes IV with the European Black--Scholes formula~\cite{alpaca_market_data_faq}, so using it in an American tree need not reproduce the quoted price. Each daily download, or snapshot, stored calls, puts, and a closing stock price in one file per ticker. We retained expirations nearest 2, 7, 14, 30, and 45 calendar days, with 60 and 90 days added on some dates. The calibration set contained fifteen snapshots labeled April 14 through May 11, 2026, with gaps between dates. Its 457 files covered 23 tickers initially and 31 thereafter: 28 stocks and three exchange-traded funds (ETFs). Folder labels, download dates, and closing-price session dates did not always agree (Supplementary Table~\ref{tab:corpus_dates}). The April 14 snapshot stored the April 13 close, and the April 29 folder stored the April 30 close; labels and sessions agreed for April 28 and May 11. We retained 234{,}549 observations with positive bids, positive calendar days to expiration (DTE), reported IV strictly between 0.01 and 2.0, and moneyness $K/S$ between 0.8 and 1.2 inclusive, where $K$ is strike and $S$ is the stored stock price. Fits used the stored DTE, which counted from the download date rather than the session date in the April 23 and May 1 snapshots. Calls and puts shared the fitting loss; repeated options contributed once per snapshot. Downloads continued through September 4, 2026, supporting sector-network refits through July 31 and scoring on later dates.

We compared a parametric surface shared by all tickers with neural surfaces shared globally, by sector, or fitted separately to individual tickers. The six sectors comprised five industry groups and one group for the three ETFs (Supplementary Table~\ref{tab:sector_map}). Every surface predicted squared IV as a ticker-specific level times a positive shape function of maturity and moneyness. Following the use of compact parametric surfaces such as SABR and SVI~\cite{hagan2002,gatheral2004}, we used a shared exponential-polynomial shape given by:
\begin{equation}\label{eq:psi}
 \psi_{\beta}(\tau,m)=\exp\!\left[
 \beta_1\ln\tau+\beta_2\ln m+\beta_3\ln\tau\ln m
 +\beta_4(\ln m)^2+\beta_5(\ln\tau)^2\right],
\end{equation}
where $\tau$ is calendar DTE in days, $m=K/S$, and $\beta_1,\ldots,\beta_5$ are fitted coefficients. For both parametric and neural shapes, the predicted squared IV for ticker $i$ was given by:
\begin{equation}\label{eq:sigma_model}
    \widehat\sigma_i^2(\tau,m)=\theta_i\,\psi(\tau,m),
    \qquad \theta_i>0,\quad \psi(\tau,m)>0,
\end{equation}
where $\theta_i$ is the ticker's variance level and $\psi$ is the shape. The parametric fit jointly estimated the shared coefficients and ticker levels using the IV loss defined below. Fits did not require total implied variance, squared IV times time to expiration, to be nondecreasing across maturities at fixed moneyness, or option prices to be convex across strikes.

The neural shape replaced the polynomial with a neural network (NN) output and was given by:
\begin{equation}\label{eq:psi_nn}
    \psi_{\mathrm{NN}}(\tau,m;w)
    =\exp\!\left[\mathrm{NN}\!\left(
    \operatorname{std}(\ln\tau),\operatorname{std}(\ln m);w\right)\right],
\end{equation}
where $w$ denotes network weights and $\operatorname{std}(\cdot)$ centers and scales each input using its training mean and standard deviation. Global and sector networks had two hidden layers of 16 $\tanh$ units ($2\to16\to16\to1$, 337 parameters). Individual tickers required at least 2{,}000 observations for their own network; those with fewer retained their sector network. Tickers with at least 5{,}000 observations used the same architecture as the global network; eligible tickers below that threshold used two layers of 8 units ($2\to8\to8\to1$, 105 parameters). We jointly fitted network weights and ticker log levels $\ln\theta_i$ by minimizing the equally weighted mean squared IV error, given by:
\begin{equation}\label{eq:nn_loss}
 \min_{w,\{\ln\theta_i\}}
 \frac{1}{n}\sum_{j=1}^{n}
 \left[
 \sqrt{\theta_{i(j)}\psi_{\mathrm{NN}}(\tau_j,m_j;w)}
 -\sigma_{\mathrm{market},j}
 \right]^2,
\end{equation}
where $n$ is the observation count, $i(j)$ identifies the ticker, and $\tau_j$, $m_j$, and $\sigma_{\mathrm{market},j}$ are observation $j$'s DTE, moneyness, and reported IV. Each log level started at twice the logarithm of its ticker's mean IV. Training used Adam with a decreasing learning rate for at most 2{,}000 epochs, or full passes through the data, and stopped after 200 epochs without training-loss improvement. We retained the final weights. A constant can shift between $\ln\theta_i$ and the network's output bias without changing predictions, so only their combined variance prediction $\theta_i\psi$ in~\eqref{eq:sigma_model} is identifiable without further normalization.

We tested transfer to withheld dates using two holdouts within the calibration window. The single-date test fitted sector networks on fourteen snapshots and tested on May 11. The April earnings test fitted them on the six snapshots labeled April 14 through April 22 and tested on April 23 and 24. For the April test, three configurations assessed the role of nearby earnings reports, with all earnings distances measured from the snapshot's folder date. Configuration A used all rows and the two inputs, moneyness and maturity. Configuration B used those inputs but excluded training and test rows within three calendar days of the ticker's or a sector peer's earnings report, testing transfer away from earnings. Configuration C used all rows and added two inputs from the public Yahoo Finance earnings calendar. The first counted calendar days to the ticker's nearest report, positive before and negative after, clipped to $[-30,30]$. The second was the smallest absolute day distance to any sector peer's report, capped at 30. For the three ETFs, both inputs used the nearest report among all 28 stocks; that distance also determined their exclusions in B. The four-input networks had 369 parameters ($4\to16\to16\to1$). The GS and LLY simulations used no earnings inputs.

\subsection{Daily IV updates}\label{sec:theta_function}

The daily IV update adapted the variance equation of the Heston model~\cite{heston1993}. In its standard form, that model pairs a stock-price diffusion with a square-root process for the stock's instantaneous variance. The variance reverts toward a long-run level, and its random changes scale with the square root of its current value. It follows the stochastic differential equation given by:
\begin{equation}\label{eq:heston_standard}
    dv_t = \kappa(\bar v-v_t)\,dt + \sigma_v\sqrt{v_t}\,dW_t^v ,
\end{equation}
where $\kappa$ is the mean-reversion rate, $\bar v$ is the long-run variance, $\sigma_v$ is the volatility of variance, and $W_t^v$ is a Brownian motion that the standard model correlates with the one driving the stock. We applied the same square-root equation to a different quantity. Each option received its own implied-variance state, which we call its IV factor. The factor is an implied variance, and its square root is the option's IV. For an option $a=(i,K,T)$ on ticker $i$ with strike $K$ and expiration $T$, the factor $v_{a,t}$ reverts toward a moving target in place of Heston's constant level $\bar v$. The target is read from the fitted surface and is given by:
\begin{equation}\label{eq:theta}
    \bar v_{a,t}=\theta_i\,
    \psi_{\mathrm{NN}}\!\left(\tau_t,K/S_t\right),
\end{equation}
where $\theta_i$ and $\psi_{\mathrm{NN}}$ are the fitted ticker level and neural shape for ticker $i$, $S_t$ is the stock price on date $t$, and $\tau_t$ is the remaining calendar DTE on that date. The target therefore moves as the stock price and the time to expiration change. The GS and LLY illustrations and the IV ablation used the per-ticker networks, and the forecast test used sector networks refitted on data through July 31. When the stock moved so far that $K/S_t$ left the fitted range of 0.8 to 1.2, the surface extrapolated, and we report how often this happened (Supplementary Table~\ref{tab:ablation_seeds}). Each factor started on the fitted surface on the entry date, April 28 in the illustrations:
\begin{equation}\label{eq:v0}
    v_{a,0}=\bar v_{a,0}.
\end{equation}
The forecast test instead started each factor at the option's observed IV on the date the forecast was issued. On every date, the IV entering the option's binomial tree was the square root of the factor:
\begin{equation}\label{eq:sigma_path}
    \sigma_{a,t}=\sqrt{v_{a,t}}.
\end{equation}
Each pricing calculation held this IV constant, and the factor changed only between trading days. The factor is therefore the implied variance of one option, not the instantaneous variance of the stock's returns.

Each factor received a random shock on every trading day, partly tied to the stock's return. We first standardized the stock's daily log returns. The standardized return shock was given by:
\begin{equation}\label{eq:return_shock}
    Z^S_{j,t}=\frac{\Delta\ln S_{j,t}-\overline{\Delta\ln S}}
    {s_{\Delta\ln S}},
\end{equation}
where $\Delta\ln S_{j,t}=\ln(S_{j,t+1}/S_{j,t})$ is the log return on path $j$ over the step from date $t$ to date $t+1$, and $\overline{\Delta\ln S}$ and $s_{\Delta\ln S}$ are a mean and standard deviation of these returns. In the GS and LLY illustrations, these two constants came from all simulated paths and dates. The IV ablation and the forecast test instead took them, together with the growth shift, from a separate pilot simulation run before the evaluation paths were drawn. No evaluation path was therefore standardized with its own later returns. The shock to each factor combined the return shock with an independent standard normal shock and was given by:
\begin{equation}\label{eq:variance_shock}
    Z^v_{j,t}=\rho Z^S_{j,t}+\sqrt{1-\rho^2}\,Z^{\perp}_{j,t},
    \qquad Z^{\perp}_{j,t}\sim\mathcal N(0,1),
\end{equation}
where $\rho$ sets how strongly the factor responds to the stock's return. We fixed $\rho=-0.6$ by assumption rather than estimating it from market data, so IV tended to rise when the stock fell. On a given path and date, every option on the same stock received the same shock $Z^v_{j,t}$, so the put and the call in each illustration responded to the same random draws. Each option still kept its own factor and target, because the surface predicts a different IV at each strike and expiration. Each factor then advanced once per trading day by an Euler step of the square-root equation, with $\Delta t=1/252$ and a floor at $v_{\min}$. The step used $Z^v_{j,t}$ in place of a normal Brownian increment, so the part of each shock tied to the stock inherited the non-normal shape of the JumpHMM returns. The floored step was given by:
\begin{equation}\label{eq:euler}
 v_{a,t+1}=\max\!\left\{
 v_{a,t}+\kappa(\bar v_{a,t}-v_{a,t})\Delta t
 +\sigma_v\sqrt{\max(v_{a,t},0)}\sqrt{\Delta t}\,Z^v_{j,t},
 v_{\min}\right\},
\end{equation}
where the path index $j$ is omitted from $v_{a,t}$ and $\bar v_{a,t}$ for brevity. We fixed $\kappa=15$ per year, $\sigma_v=0.5$, and $v_{\min}=(0.005)^2$ as assumptions rather than estimates. With daily steps, $\kappa=15$ closes about 6\% of the gap between a factor and its target each trading day. The floor kept the factor positive for the pricer. It was needed because an Euler step can fall below zero even when the Feller condition, which keeps the continuous-time process positive, holds~\cite{andersen2008}. The simulation code also contains two optional features, a target that depends on JumpHMM's hidden state and a multiplier that raises the target during visits to tail states. Both were switched off in every analysis, so the fitted surface alone set the target and $\rho$ alone linked the random shocks to stock returns.

\subsection{American-option pricing and profit and loss}

A binomial tree prices an American option by working backward from expiration and taking, at every node, the larger of the value of continuing and the value of exercising immediately~\cite{cox1979,broadie1996}. We used two such trees, each with one constant volatility at every node. The Cox--Ross--Rubinstein (CRR) tree sets its up and down moves from that volatility~\cite{cox1979}. The Leisen--Reimer (LR) tree uses the Peizer--Pratt inversion to place its moves so that prices converge faster as steps are added, which gives the smooth prices that finite-difference Greeks need~\cite{leisen1996}. Moment-matching trees such as those of Jarrow--Rudd and Tian~\cite{jarrow1983,tian1993} are further alternatives that we did not use. To turn fitted IVs into dollar prices for comparison with the quoted midpoints in the first analysis, we used a 200-step American CRR tree. All simulated prices and Greeks used a 201-step American LR tree, since the LR construction requires an odd number of steps. When the option's IV $\sigma_{a,t}$ was so small that $\sigma_{a,t}\sqrt{\tau_t/365}<0.01$, with $\tau_t/365$ the time to expiration in years, or the LR tree gave no finite price, an American CRR tree with at least as many steps priced the option instead. Both trees priced options risk-neutrally with an assumed constant risk-free rate of $4.25\%$. This rate is separate from the zero rate that defines JumpHMM's growth rates, and the simulated stock paths followed JumpHMM's real-world growth rates rather than the risk-free rate. Both also set the dividend yield to zero ($q=0$) for every ticker, although GS, LLY, and many other tickers pay dividends. We accepted both mismatches as simplifications. With no dividends and a positive rate, early exercise of an American call is never optimal, so the American feature mattered only for the puts.

The GS and LLY illustrations each simulated one put and one call, opened on April 28, 2026, and expiring on May 29, 2026. On April 28, GS closed at \$926.38, and its options had strikes of \$890 (put, $K/S=0.961$) and \$970 (call, $K/S=1.047$). For LLY, the close was \$873.83, and the options had strikes of \$825 (put, $K/S=0.944$) and \$940 (call, $K/S=1.076$). All four options were therefore out of the money. We modeled short positions entered at the indicative midpoints in the April 28 snapshot, whose label, download, and session dates all agree. The modeled cost of buying back an option can exceed its entry premium even if it later expires worthless. The holding period covered 31 calendar days and 22 trading days after excluding weekends and the May 25 market holiday. The tree measured time to expiration in calendar days and received $\tau_t/365$ years, while the stock and the factors advanced in 22 daily steps of $\Delta t=1/252$. The illustrations used 1{,}000 paths per ticker with seed 20260429. Each path started at the April 28 closing price, with its hidden state drawn from the stationary distribution of the transition matrix. Its 22 simulated returns then gave the prices on the 22 later trading days. Because the calibration window ran to May 11, the surfaces used in the illustrations and the IV ablation included quotes from after April 28, so these analyses illustrate the model rather than forecast. At expiration, the option's intrinsic value, its payoff from immediate exercise, replaced the tree price (Algorithm~\ref{alg:forward_sim}).

\begin{algorithm}[H]
\caption{Forward simulation of contract values under the full factor; the other IV variants change only the factor update}\label{alg:forward_sim}
\begin{algorithmic}[1]
\Require JumpHMM model for ticker $i$ and growth shift $c$; fitted $(\theta_i,\psi_{\mathrm{NN}})$; contracts $a=(K,T,\phi)$, with $\phi=1$ for a call and $\phi=-1$ for a put; trading dates $d_0,\ldots,d_L$; calendar DTE $\tau_0,\ldots,\tau_L$; $(\kappa,\sigma_v,\rho)$; entry price $S_0$; risk-free rate $r$; $N_p$ paths; LR tree steps $N_{\mathrm{LR}}$.
\Ensure Stock, IV-factor, IV, and option-value paths.
\State Draw $N_p$ JumpHMM stock paths with $L$ trading transitions and emissions truncated as in~\eqref{eq:truncated_emission}.
\State Compute $Z^S_{j,\ell}$ by standardizing the simulated daily log returns as in~\eqref{eq:return_shock}, with constants from these paths or from a pilot simulation.
\State Draw $Z^\perp_{j,\ell}\sim\mathcal N(0,1)$ once per path and transition; set $Z^v_{j,\ell}\gets\rho Z^S_{j,\ell}+\sqrt{1-\rho^2}Z^\perp_{j,\ell}$ and reuse it for every contract.
\For{contract $a$ and path $j$}
  \State $v_{a,0}^{(j)} \gets \theta_i\psi_{\mathrm{NN}}(\tau_0,K/S_0)$
  \For{$\ell=0$ \textbf{to} $L-1$}
    \State $\sigma_{a,\ell}^{(j)} \gets \sqrt{v_{a,\ell}^{(j)}}$
    \State $P_{a,\ell}^{(j)} \gets \mathrm{LR}_{N_{\mathrm{LR}}}(S_{j,\ell},K,\sigma_{a,\ell}^{(j)},\tau_\ell/365,r)$
    \State $\bar v_{a,\ell}^{(j)}\gets\theta_i\psi_{\mathrm{NN}}(\tau_\ell,K/S_{j,\ell})$
    \State Update $v_{a,\ell+1}^{(j)}$ with the floored Euler step in~\eqref{eq:euler} using $\Delta t=1/252$.
  \EndFor
  \State $P_{a,L}^{(j)}\gets\max\{\phi(S_{j,L}-K),0\}$.
\EndFor
\State \Return all paths.
\end{algorithmic}
\end{algorithm}

For a short position sold at premium $p_0$, profit and loss (P\&L) at expiration was given by:
\begin{equation}\label{eq:terminal_pnl}
 \Pi_T^{\mathrm{put}}=p_0^{\mathrm{put}}-\max(K-S_T,0),
 \qquad
 \Pi_T^{\mathrm{call}}=p_0^{\mathrm{call}}-\max(S_T-K,0),
\end{equation}
where $S_T$ is the JumpHMM stock price at expiration. We assumed entry at the observed midpoint, no transaction costs, no margin constraint, and no interest on the premium received. We also ignored early assignment. We valued each short position at expiration and, in the IV ablation, after five, ten, fifteen, or twenty trading days. These valuation dates were fixed across paths, even though the tree prices include the holder's right to exercise early. With early assignment ignored, the P\&L at expiration depends only on $S_T$, the strike, and the premium, so the IV updates changed option prices and Greeks before expiration but not the payoff of a position held to expiration. All premiums, option values, and P\&L are in dollars per share, so amounts for a standard 100-share contract are 100 times these values. We calculated the Greeks on the same LR tree by central finite differences, which reprice the option after equal upward and downward changes to one input. Delta measures the price's sensitivity to the stock price, gamma measures how delta changes with the stock price, and Vega measures the price's sensitivity to IV. Delta and gamma used stock-price bumps of $\pm1.5\%$. Vega used IV bumps of $\pm0.005$ in absolute terms and is reported per volatility percentage point.

\subsection{Comparison of five IV variants}\label{sec:ablation_methods}

To find which parts of the IV update changed option values before expiration, we repriced the GS and LLY options, sold at the same April 28 premiums, under five IV variants in the IV ablation. Within each seed, the five variants shared the same simulated stock paths. The first ablation seed equals the illustration seed, but the ablation set its growth shift from the pilot simulation, so its paths are not the illustration paths. Frozen IV kept each option's IV at its fitted entry value. The direct surface set the factor equal to its target, $v_{a,t}=\bar v_{a,t}$, on every date. Mean reversion, a deterministic variant, used the update in~\eqref{eq:euler} without shocks ($\sigma_v=0$). The uncoupled factor added random shocks unrelated to stock returns ($\sigma_v=0.5$, $\rho=0$). The full factor linked those shocks to stock returns ($\sigma_v=0.5$, $\rho=-0.6$). All three variants that followed~\eqref{eq:euler} used $\kappa=15$. Every variant started from the same fitted entry IV, and the two stochastic variants reused the same independent normal shocks, so results could be compared path by path. We ran the ablation with three random-number seeds per ticker (the integers 20260429, 20260430, and 20260431) and 1{,}000 paths per seed. Its pilot simulation used 10{,}000 paths (seed 20260905) and the same 22 trading-day steps. It fixed the growth shift for the 10\% target and the return mean and standard deviation before the evaluation paths were drawn.

We compared the variants by hypothetical short-position P\&L at fixed horizons. The primary horizon was ten trading days, on May 12 with 17 calendar days to expiration; secondary horizons were five, fifteen, and twenty trading days. We used no profit target, stop-loss, or rule for choosing when to close a position. We used the direct surface as the baseline, because its IV sits exactly on the target that the three updating variants move toward. The hypothetical P\&L at each horizon, and its difference from the direct surface on the same path, were given by:
\begin{equation}\label{eq:ablation_pnl}
    \Pi_{a,t}^{(g,j)}=p_{a,0}-P_{a,t}^{(g,j)},
    \qquad D_{a,t}^{(g,j)}=\Pi_{a,t}^{(g,j)}-\Pi_{a,t}^{(\mathrm{surface},j)},
\end{equation}
where $g$ denotes the IV variant, $j$ the shared stock path, $p_{a,0}$ the entry premium, and $P_{a,t}^{(g,j)}$ the option's tree price on date $t$. Because the premium is the market midpoint while later prices come from the tree, $\Pi$ also includes the gap between the two at entry, and this gap cancels in $D$. For each variant, we reported the mean of $\Pi$, its standard deviation, its 5\% quantile, and its 5\% expected shortfall, defined as the mean P\&L at or below that quantile. The Monte Carlo standard error of the mean paired difference was $s_D/\sqrt{N_p}$, where $s_D$ is the standard deviation of $D$ across the $N_p$ paths, 1{,}000 per seed or 3{,}000 pooled. These standard errors measure simulation noise with the fitted models and pilot constants held fixed. Price differences on individual paths can cancel in the mean, so we also reported the mean absolute difference between each variant's price and the direct-surface price on the same path. We reported each seed separately and also pooled the 3{,}000 paths per ticker.

We checked that entry prices and payoffs at expiration agreed exactly across variants, as they must by construction. We also checked the numerical accuracy of the prices in between. For every variant, we compared 201-step and 401-step LR prices on the first twenty paths of every seed after ten and twenty trading days and recorded the largest absolute difference. The tree prices each option separately, so nothing forces prices to be consistent across strikes. We therefore priced calls and puts at eleven fixed strikes from 0.8 to 1.2 times the entry stock price, a check we call the strike audit. The audit ran within the first ablation seed. Each audit strike used the May 29 expiration and its own factor and target, driven by the same shocks as the GS or LLY options. The audit used the first fifty paths of the first seed at entry and after ten and twenty trading days, at both 201 and 401 steps. Entry, which all paths share, was checked once. Each ticker, variant, option type, and tree size therefore had 101 stock-price points, the entry price and fifty paths on each of two dates. At each of these points, we checked four conditions, and a violation's size was the dollar amount by which a condition failed. Each price had to lie between its intrinsic value and an upper bound, the stock price for a call and the strike for a put. As the strike increased, call prices could not rise and put prices could not fall. The price difference between adjacent strikes could not exceed the distance between the strikes (the vertical-spread bound). Prices had to be convex in strike, so no price could exceed the straight line between its neighbors. For all four checks, we counted violations larger than \$0.01 per share. Passing these checks is necessary, but not sufficient, for prices to be free of arbitrage across strikes. The audit compared strikes within one expiration on each date, so it did not test consistency across expiration dates.

\subsection{Tests on later dates}\label{sec:chronological_methods}

We tested the sector surfaces on later dates before using them to forecast option prices. The first stage used 58 snapshots labeled April 14 through July 17, 2026, in 53 expanding-window folds. Each fold refitted the networks on earlier snapshots and predicted IV on the next, using that date's observed stock price to compute moneyness. Its root mean squared error (RMSE) therefore measured surface prediction without a stock forecast or an IV update. The second stage fitted six sector networks on 68 sessions through July 31 and fixed them before testing on 23 sessions from August 4 through September 4. We scored IV using observed test-date stock prices, reporting RMSE and bias, the mean predicted minus observed IV. Unlike the first stage, which used files as saved, this stage audited retrieval times, excluded intraday downloads, required two-sided quotes, and computed DTE from the stored close's session date. Training excluded files retrieved after July 31. Retrieval times could not establish quote freshness or exact stock--option synchronization, so timing differences could affect both IV and price errors. Supplementary Section~\ref{sec:chronological_supp} gives the data and fitting rules and the exploratory earnings analysis performed after the July-cutoff surface test. Neither stage tested the per-ticker networks used in the April illustrations.

The option test issued forecasts at successive observed sessions, called origins, from August 4 into early September. At each origin, we selected one put and one call per ticker (GS and LLY) in each of two maturity cohorts using only origin information. Monthly options had 25--45 calendar DTE and were scored one trading session later; shorter-maturity options had 10--21 DTE and were scored one and five trading sessions later. The shorter cohort was added after none of the monthly options remained in the downloaded maturity ladder at five sessions, with this protocol amendment recorded before inspecting aggregate forecast errors. Only exact-contract quotes at the specified endpoint were scored. Forecasts used observed origin stock prices and IVs, with the July-cutoff surfaces, historical JumpHMM parameters, and IV-update constants fixed. All five IV variants shared 1{,}000 stock paths and shocks per origin. Stock states started from the stationary distribution, without conditioning on recent returns, and drift and shock normalization came from an independent pilot. To remove the fitted surface's initial IV-level error from the comparison, we rescaled its variance target at each origin as follows:
\begin{equation}\label{eq:origin_anchor}
\bar v^{*}_{a,t}=\sigma_{a,0,\mathrm{obs}}^2
\frac{\bar v_{a,t}}{\bar v_{a,0}},
\end{equation}
where $\sigma_{a,0,\mathrm{obs}}$ is observed origin IV and $\bar v_{a,t}$ is the surface target in~\eqref{eq:theta}. Frozen IV held the observed value, the direct surface followed $\bar v^{*}_{a,t}$, and the three updating variants reverted toward it with the IV-ablation parameters. This anchoring applied only to forecasts; the ablation retained its fitted starting values.

We fixed the primary score before evaluation: the mean absolute error (MAE) of the forecast mean against the later bid--ask midpoint. We also scored forecast medians and reported signed error, RMSE, and the fraction of forecast means inside the quoted spread. Distributional scores comprised the continuous ranked probability score (CRPS), which rewards concentration near the observed value, and central 90\% interval coverage and width. The CRPS definition and simulation settings appear in Supplementary Section~\ref{sec:chronological_supp}. All monetary scores were in dollars per share. Scores weighted origins equally after averaging matched-contract contributions within each ticker, cohort, horizon, and origin. Overlapping horizons and shared market moves made origins dependent; more simulated paths did not increase the market sample size. To assess the contribution of stock-forecast error, we repeated IV updates along observed stock paths when every intervening session was available, retaining the independent IV shocks. A second diagnostic repriced options in the same tree using observed endpoint stock prices and reported IVs to measure the tree's departure from quoted midpoints. Both used future observations and were diagnostics, not forecasts. We also compared JumpHMM stock forecasts with a zero-log-drift Gaussian random walk. Supplementary diagnostics examined simulated IV changes, stock drift and volatility, and earnings-related surface bias.

\subsection{Stock forecast benchmarks}\label{sec:stock_validation_methods}

We investigated JumpHMM's stock-forecast errors with separate 10{,}000-path runs, leaving the option test's 1{,}000-path forecasts and the IV-ablation paths unchanged. We checked the conversion of annualized growth rates to daily log returns with the $1/252$ step and daily compounding by reconstructing prices from saved growth rates. An initialization diagnostic repeated the 2026 forecasts with states inferred from available past returns. Its forward filter kept the fitted parameters and tracked both the hidden state and remaining jump duration using the truncated-emission likelihoods in~\eqref{eq:truncated_emission}. The four-method benchmark instead retained JumpHMM's stationary initialization and pilot growth shift, comparing it with unchanged-price, adaptive-volatility, and direction-predicting forecasts. The forecast test's random walk and filtered initialization remained separate diagnostics. The unchanged-price benchmark predicted the origin price without an uncertainty interval. The adaptive-volatility model used zero-drift Gaussian log returns with variance estimated by an exponentially weighted moving average (EWMA)~\cite{riskmetrics1996}; the direction-predicting model added a regression-based drift. Their variance update and forecast return were given by:
\begin{equation}\label{eq:stock_benchmark}
a_t=\lambda a_{t-1}+(1-\lambda)u_t^2,
\qquad
\log\frac{S_{t+h}}{S_t}=h\mu_t+\sqrt{a_t}\sum_{j=1}^{h}Z_j,
\quad Z_j\sim\mathcal N(0,1),
\end{equation}
where $u_t=\log(S_t/S_{t-1})$ is the observed daily log return, $a_t$ is daily variance estimated at origin $t$, $\lambda$ is its decay, and $h$ is the horizon in trading days. The adaptive-volatility model set daily drift $\mu_t=0$; the direction-predicting model used the regression below. Both held drift and variance fixed within the horizon. Only JumpHMM used the return cutoff.

The direction regression predicted the next daily log return divided by origin EWMA volatility; multiplying its prediction by that volatility gave $\mu_t$. Its four inputs were the stock's cumulative log returns over its latest one, five, and twenty daily returns, and SPY's cumulative log return over its latest five daily returns. The fund SPY tracks the S\&P 500. Each was divided by the square root of its window length and its own asset's origin EWMA volatility, then by that input's training root mean square. We fitted GS and LLY separately without intercepts, using ridge penalties to shrink coefficients toward zero~\cite{hoerl1970}. Annual validation blocks from 2019 through 2024 used only earlier data to fit coefficients and input scales. We selected one decay and one penalty for both tickers, weighting ticker, validation year, and horizon (one and five trading sessions) equally. Candidate decays were $0.80$, $0.90$, $0.94$, $0.97$, and $0.99$; selection minimized Gaussian log loss of cumulative log returns under zero drift. Candidate penalties ranged from $0.01$ to $1000$, plus the zero-drift limit of an infinite penalty, which sets all coefficients to zero. With the selected decay fixed, penalty selection minimized squared cumulative-return error divided by forecast variance. We then fitted coefficients and input scales on 2014--2024 observations and froze them.

We tested all four methods on 2025 and August--September 2026 data, using identical origin prices and endpoints. Historical closes came from the code's S\&P 500 daily price files; 2026 closes came from the option files. The latter began in April, when the EWMA variance restarted from its 2014--2024 estimate. The 2025 test had 249 one-session and 245 five-session origins per ticker; the 2026 test used option-test origins with observed endpoints, giving 21 and 17, respectively. Origin features and variance used only available past data. We computed daily returns only from closes on adjacent trading sessions, omitting returns across gaps. Each stochastic method used 10{,}000 paths per origin under three fixed seeds, with Gaussian shocks shared by the adaptive-volatility and direction-predicting models. We scored median forecasts by MAE and mean forecasts by RMSE, matching each summary to the loss it minimizes~\cite{hyndman2021}; the option test retained its prespecified mean-forecast MAE. For the three stochastic methods, distributional scores were CRPS and central 90\% interval coverage and width. Scores averaged dates within each seed, then seeds. To interpret interval widths, we compared JumpHMM's pilot daily log-return standard deviation with the observed test-period value and the adaptive-volatility estimate averaged over origins (Supplementary Table~\ref{tab:mechanism_checks}). Although settings came from 2019--2024 validation, we proposed the alternative models after seeing the 2026 errors. We therefore reported the periods separately and treated the 2026 comparison as exploratory.

\section{Results}\label{sec:results}

\begin{figure}[H]
\centering
\includegraphics[width=\textwidth]{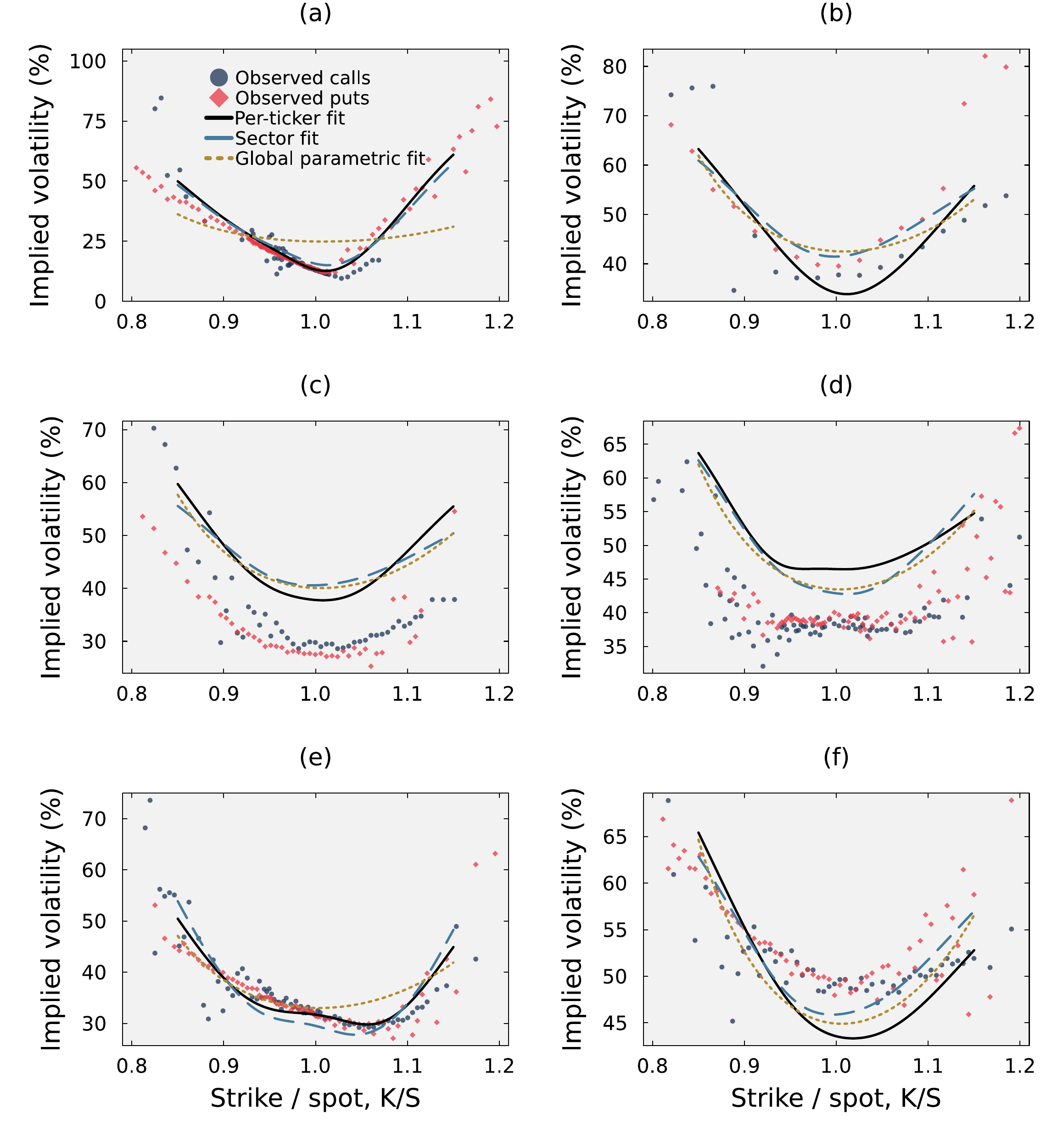}
\caption{In-sample comparison of the implied volatility (IV) observed on May 11, 2026 with surfaces fitted across all fifteen calibration snapshots. Panels show (a) SPY, (b) NVDA, (c) MSFT, (d) LLY, (e) GS, and (f) AVGO, with 7, 7, 9, 11, 11, and 9 calendar days to expiration, respectively. Circles show reported call IV, and diamonds show reported put IV. Solid, dashed, and dotted curves show per-ticker neural, sector neural, and global parametric fits. The legend applies to all panels. The per-ticker fits for MSFT and LLY overpredict IV near the money (Supplementary Table~\ref{tab:smile_date_bias}).}\label{fig:per_ticker_panels}
\end{figure}

\begin{figure}[H]
\centering
\includegraphics[width=\textwidth]{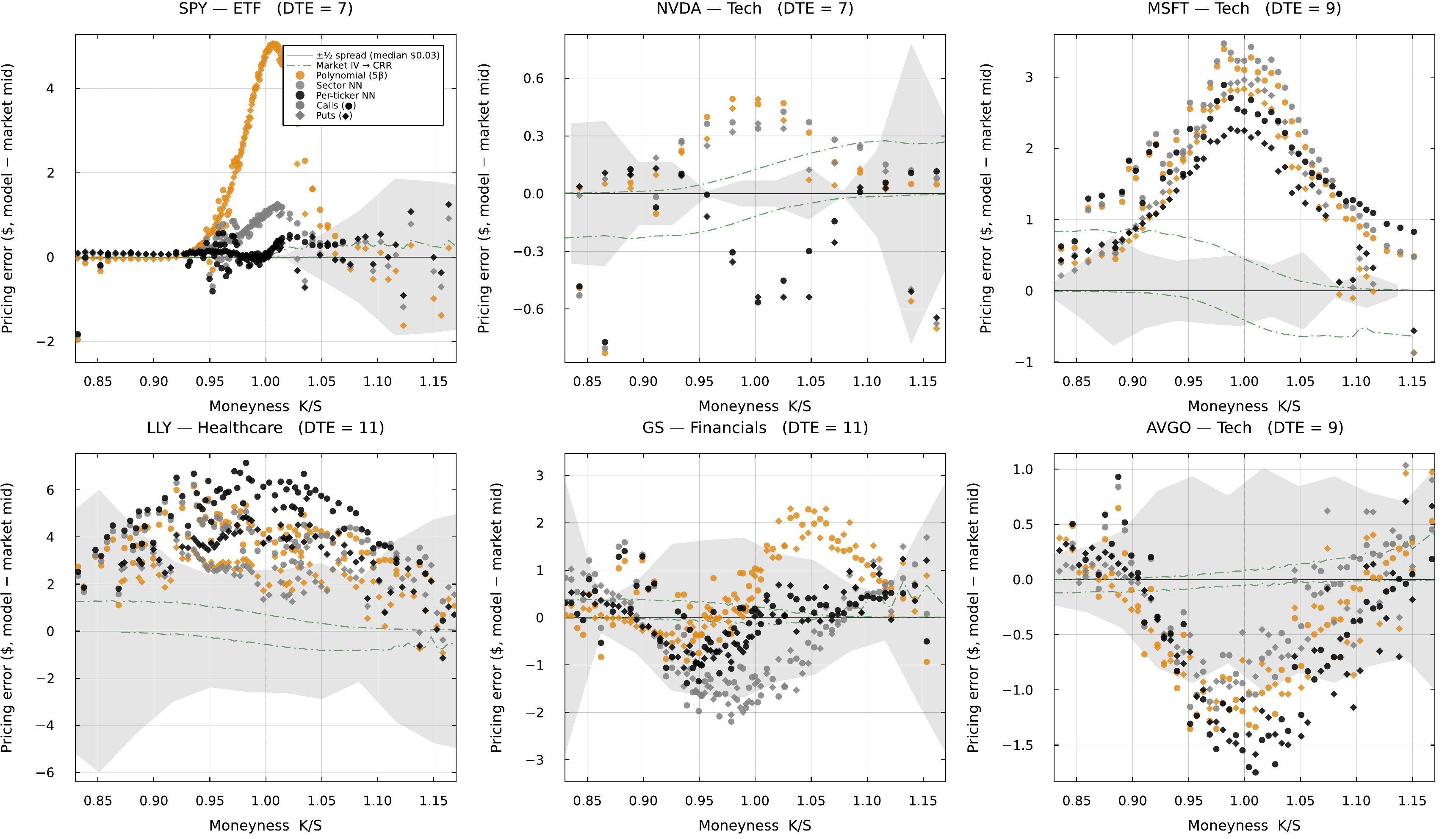}
\caption{In-sample comparison of CRR American-option prices with Alpaca indicative mids across moneyness for SPY, NVDA, MSFT, LLY, GS, and AVGO on May 11, 2026, at the maturities of Fig.~\ref{fig:per_ticker_panels}. Errors are tree price minus indicative mid. Points show errors of prices from the parametric, sector, and per-ticker surfaces, with circles for calls and diamonds for puts. Dash-dotted curves show, for reference, errors of prices from each contract's reported IV. The gray ribbon spans plus and minus the median bid--ask half-spread in each moneyness bin. A point inside the ribbon has an absolute pricing error below the typical half-spread at that moneyness. The shares reported in the text instead compare each contract with its own half-spread.}\label{fig:price_error}
\end{figure}

More flexible IV surfaces lowered pooled fitting error, with the per-ticker models giving the closest pooled fit (Table~\ref{tab:three_way}; Figs.~\ref{fig:per_ticker_panels} and~\ref{fig:price_error}). Across 234{,}549 observations from 31 tickers, root mean squared error (RMSE), measured in volatility percentage points, fell from 12.48 for the global parametric surface to 11.47 for a global neural surface and 10.24 for sector networks. Per-ticker networks, with sector fallbacks for PFE and BMY, reduced it to 9.73 points, 22\% below the parametric fit. Of the 29 tickers with their own network, 21 improved on their sector fit; the eight exceptions used smaller networks (Supplementary Table~\ref{tab:per_ticker_comparison}). These pooled gains did not remove errors within the calibration period. On May 11, the per-ticker fits overpredicted near-the-money IV by 9.45 points for MSFT and 7.90 for LLY, while underpredicting AVGO by 5.37 points (Supplementary Table~\ref{tab:smile_date_bias}). Each surface had one ticker level across all fifteen snapshots and could not shift it for a particular date. Pricing the six illustrated tickers with their per-ticker fits showed why dollar errors also mattered. Only 20\% of MSFT absolute price errors were within the contract's bid--ask half-spread, compared with 88\% for GS (Fig.~\ref{fig:price_error}; Supplementary Table~\ref{tab:price_error_summary}). These shares also depended on spread width. For example, only about one third of SPY prices fell inside their spreads despite small IV bias; its median half-spread was \$0.03, compared with \$1.56 for GS. For LLY, replacing fitted IV with reported IV in the same tree cut median absolute price error from \$4.01 to \$0.67 per share. Thus, even an in-sample surface error could materially affect option prices.

\begin{table}[H]
\centering
\caption{In-sample IV RMSE of parametric and neural surfaces on 234{,}549 observations, in volatility percentage points. Each model fitted a separate level for each ticker jointly with its shape parameters. The final row used per-ticker neural surfaces for the 29 tickers with at least 2{,}000 observations and sector surfaces for PFE and BMY, which had fewer.}\label{tab:three_way}
\begin{tabular}{lrr}
\toprule
Model & RMSE & Reduction from parametric \\
\midrule
Parametric shape, global               & 12.48 & reference \\
Neural shape, global                   & 11.47 & 8.1\% \\
Neural shape, six sectors              & 10.24 & 17.9\% \\
Neural shape, per-ticker with fallback &  9.73 & 22.0\% \\
\bottomrule
\end{tabular}
\end{table}

Adding earnings-date inputs reduced sector-surface prediction errors in the April holdout (Table~\ref{tab:earnings_holdout}). With May 11 excluded from the sector fits, pooled RMSE was 10.21 points in training and 10.89 on that date, with the largest increases in energy and retail (Supplementary Table~\ref{tab:loo_sector_nn}). The April earnings holdout produced a larger gap: configuration A, using moneyness and maturity alone, rose from 8.05 to 13.03 points. Removing observations near earnings in configuration B removed the gap, but changed both samples and discarded most rows, so it did not establish that earnings caused the gap in A. Configuration C used the same rows as A and added earnings-date inputs, reducing test RMSE to 11.98 points. Improvements varied across technology tickers (Supplementary Table~\ref{tab:tech_pivot}). Correcting the one-day DTE offset in the April 23 snapshot raised test RMSE by 0.14--0.19 points without changing these comparisons (Table~\ref{tab:earnings_holdout}). Earnings inputs therefore reduced the transfer error only partly. These holdouts tested sector networks, not the per-ticker surfaces used in the GS and LLY illustrations.

\begin{table}[H]
\centering
\caption{April earnings-holdout errors for three configurations of the sector networks. Training used six snapshots labeled April 14 through April 22, 2026, and testing used April 23--24. Every test ticker appeared in training. Errors and gaps are in volatility percentage points, and each gap is test minus training RMSE, computed before rounding. Configurations A and C used the same rows and differed only in their inputs. Configuration B removed rows within three calendar days of the ticker's or a peer's earnings date. The columns $N_{\mathrm{tr}}$ and $N_{\mathrm{te}}$ count training and test rows. Test rows from the April 23 snapshot used the stored DTE, which counted from the April 24 download. Counting them from the April 23 session instead, with the same trained networks, gave test RMSEs of 13.22, 8.10, and 12.15 for Configurations A, B, and C.}\label{tab:earnings_holdout}
\resizebox{\textwidth}{!}{%
\begin{tabular}{lrrrrr}
\toprule
Configuration & $N_{\mathrm{tr}}$ & $N_{\mathrm{te}}$ & Train RMSE & Test RMSE & Gap \\
\midrule
A: pooled, two inputs          & 103{,}897 & 26{,}833 & 8.05 & 13.03 & $+4.99$ \\
B: non-earnings, two inputs    &  29{,}383 &  4{,}516 & 8.02 &  7.96 & $-0.06$ \\
C: earnings-aware, four inputs & 103{,}897 & 26{,}833 & 7.61 & 11.98 & $+4.36$ \\
\bottomrule
\end{tabular}}
\end{table}

\begin{figure}[!htbp]
\centering
\includegraphics[width=\textwidth]{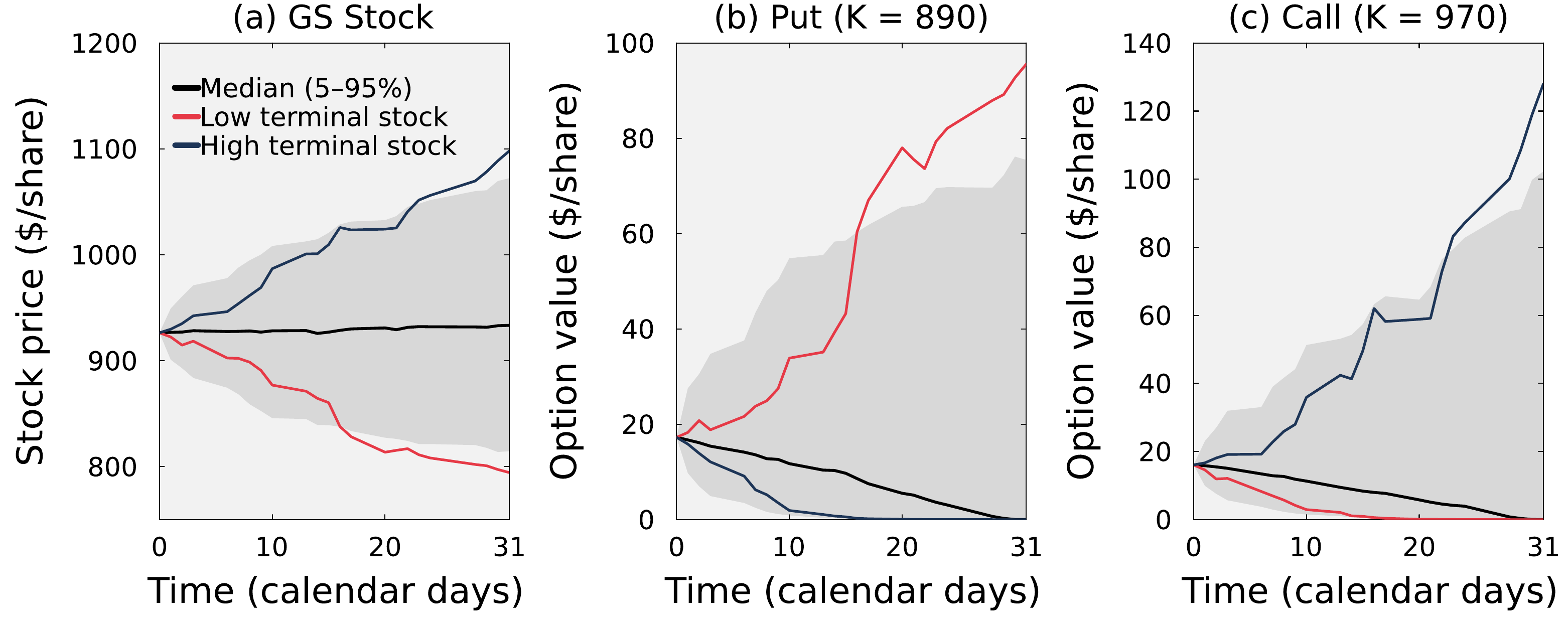}
\caption{Simulated evolution of GS stock prices (a), put values (b), and call values (c) along 1{,}000 paths between April 28 and May 29, 2026. Time is measured in calendar days from April 28. Black curves show the median at each date, and shaded bands contain the middle 90\% of values. Red curves show the median of the paths whose terminal stock price ranked in the 1st--5th percentiles, and blue curves show the median of those in the 95th--99th percentiles. Option values are dollars per share and represent the modeled cost of repurchasing the contract for a short seller. Entry premiums were \$16.51 per share for the put and \$16.09 for the call.}
\label{fig:gs_paths}
\end{figure}

\begin{figure}[!htbp]
\centering
\includegraphics[width=\textwidth]{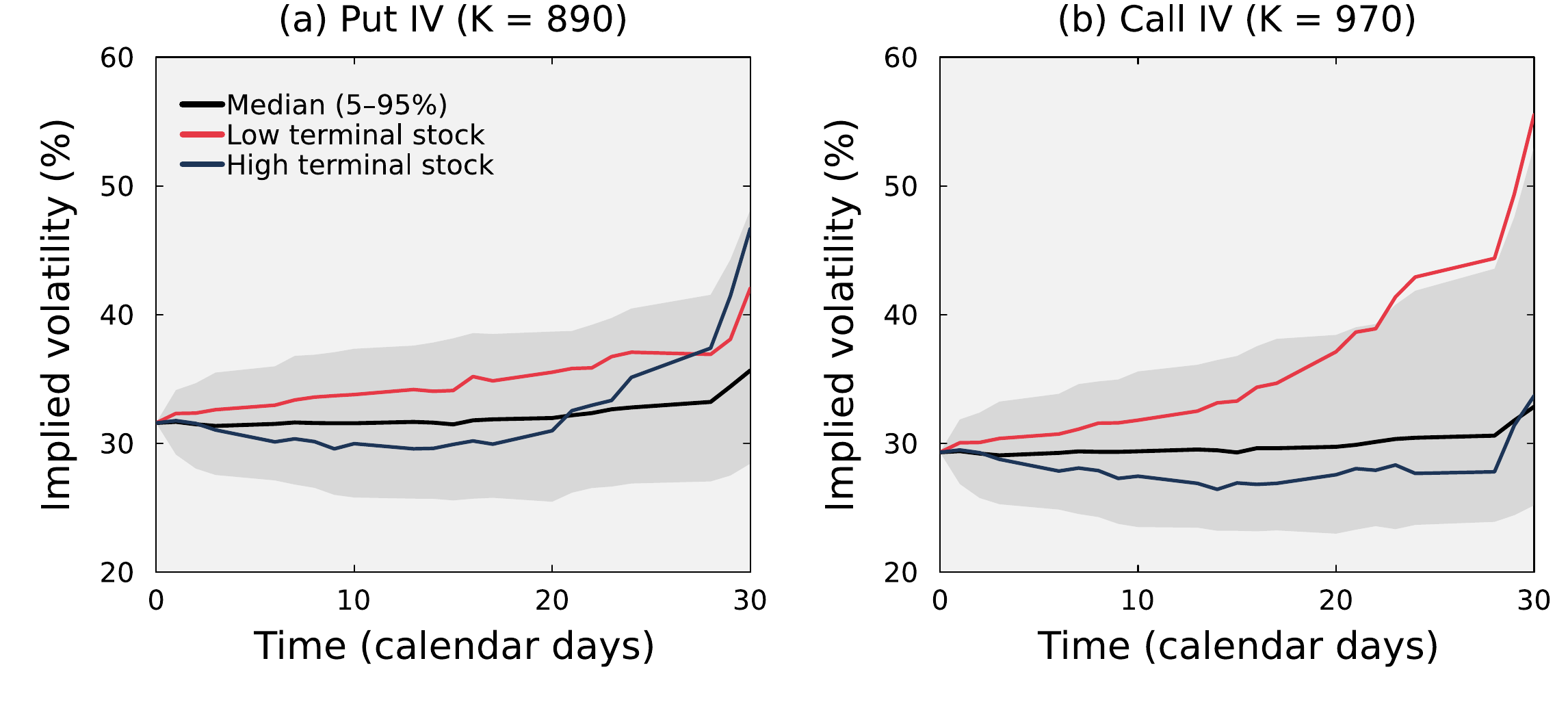}
\caption{Simulated evolution of the annualized IV of the GS put (a) and call (b) in the scenario of Fig.~\ref{fig:gs_paths}. Both panels express IV as percentages on the same vertical scale. Curves show pointwise medians across all 1{,}000 paths and within the terminal-stock groups defined in Fig.~\ref{fig:gs_paths}. Shading shows the 5th--95th percentile band across all 1{,}000 paths. Time is measured in calendar days from April 28, 2026, and stops before expiration because the payoff at expiration does not depend on IV.}
\label{fig:gs_iv}
\end{figure}

\FloatBarrier
The GS and LLY illustrations showed how contract-specific IV targets changed option values along fixed stock paths (Figs.~\ref{fig:gs_paths} and~\ref{fig:gs_iv}; Supplementary Figs.~\ref{fig:lly_paths} and~\ref{fig:lly_iv}). Each ticker used one put and one call from April 28, expiring May 29, and 1{,}000 paths with the shared settings in Supplementary Table~\ref{tab:scenario_settings}. On the lowest-terminal-price GS paths, both median IVs rose as the stock fell. The call's higher strike gave it higher moneyness and a higher target on the surface's rising right wing, so its IV rose more. Part of the late rise occurred beyond the fitted moneyness limit of 1.2. Targets also rose as expiration approached, lifting both IVs near the end. Near expiration, delta approached $\pm1$ on paths heading toward exercise, gamma concentrated near the strike, and Vega declined (Supplementary Figs.~\ref{fig:gs_greeks} and~\ref{fig:lly_greeks}). These were illustrations, not forecasts, because the surfaces included later quotes. At expiration, GS short puts and calls retained their full premiums on 74.0\% and 67.7\% of paths, but the worst 5\% lost at least \$58.96 and \$86.18 per share. For LLY, the corresponding premium-retention frequencies were 80.7\% and 79.3\% (Supplementary Tables~\ref{tab:gs_scenario} and~\ref{tab:lly_scenario}). With early assignment ignored, these terminal outcomes depended on stock prices, strikes, and premiums, not the IV update. Its effect therefore had to be measured before expiration.

\begin{figure}[H]
\centering
\includegraphics[width=\textwidth]{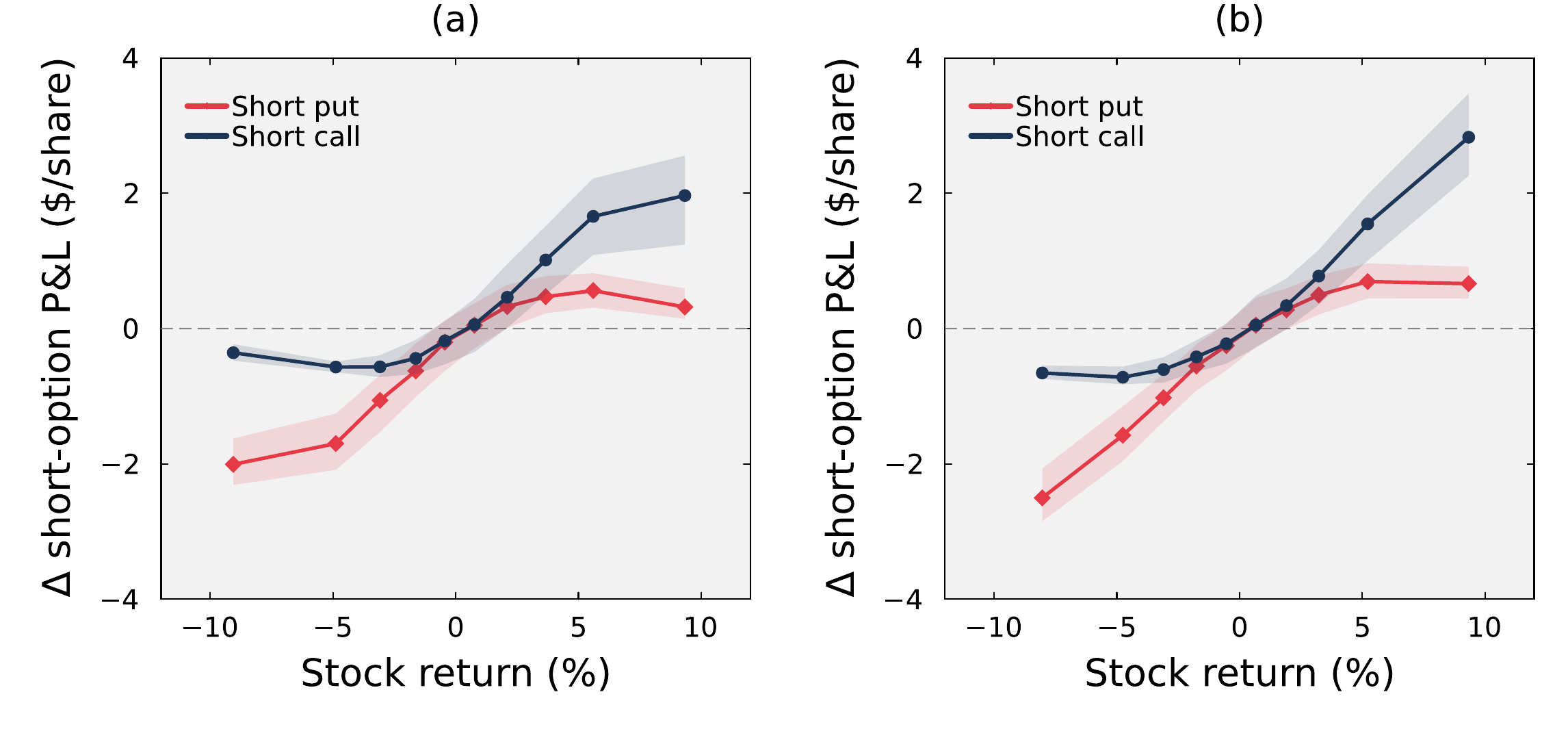}
\caption{Change in simulated short-option P\&L at the fixed horizon of ten trading days caused by linking IV shocks to stock returns, for (a) GS and (b) LLY. Each difference is full-factor P\&L ($\rho=-0.6$) minus uncoupled-factor P\&L ($\rho=0$) on the same stock path, with the same independent normal shocks. Negative values indicate worse short-position P\&L under coupling. The horizontal axis is each path's stock return from entry on April 28 to May 12. The 3{,}000 paths per ticker are sorted by this return and divided into ten groups of 300. Symbols show each group's median P\&L difference at its median stock return, and lines connect the symbols. Shaded bands span each group's 25th--75th percentiles. These bands describe variation across paths, not confidence intervals. Red diamonds denote puts and navy circles denote calls, as in Figs.~\ref{fig:per_ticker_panels} and~\ref{fig:price_error}. All prices are evaluated on May 12, 2026, with 17 calendar days to expiration.}\label{fig:dynamic_ablation}
\end{figure}

\FloatBarrier
Repricing the same four options on identical stock paths isolated the effect of the IV rule before expiration (Supplementary Table~\ref{tab:dynamic_ablation}). After ten trading days, across 3{,}000 paths, the full factor's mean absolute price difference from the direct surface was \$1.77--\$2.41 per share, compared with \$0.51--\$2.23 for frozen IV and \$0.42--\$1.82 for mean reversion. Similar differences across three seeds showed that lag alone changed prices, while random shocks added larger pathwise changes (Supplementary Table~\ref{tab:ablation_seeds}). Mean P\&L changed less because positive and negative differences partly canceled. Linking IV shocks to stock returns, or coupling, then shifted short-position P\&L from falling-stock paths toward rising-stock paths (Fig.~\ref{fig:dynamic_ablation}). Relative to the uncoupled factor, grouped path comparisons showed deeper put losses when stocks fell and smaller call losses when stocks rose. For GS, coupling changed the 5\% expected shortfall, the mean P\&L in the worst 5\% of outcomes, from $-\$62.98$ to $-\$64.60$ for the put and from $-\$71.53$ to $-\$69.82$ for the call. The LLY shifts had the same signs, and all four were at least seven times their paired-bootstrap simulation standard errors (Supplementary Table~\ref{tab:dynamic_ablation}). Entry prices and terminal payoffs were identical across variants, so an expiration-only comparison would have missed these effects.

The differences between IV variants were much larger than the price changes under tree refinement. Increasing the tree from 201 to 401 steps changed sampled prices by at most \$0.0017 per share; no factor reached its variance floor, and only 0.7--2.9\% of simulated dates for the four illustrated options left the fitted moneyness range (Supplementary Table~\ref{tab:ablation_seeds}). In the strike audit, the GS direct surface failed 178 of 1{,}818 convexity checks by more than \$0.01, with a largest excess of \$0.113 per share. These failures implied negative-priced butterfly spreads. It also failed 68 monotonicity and 68 vertical-spread checks, on different paths and dates (Supplementary Table~\ref{tab:strike_audit}). Violations occurred after entry as stock prices and maturities changed, including where audit strikes required extrapolation. The updating variants had no convexity violations and at most three monotonicity violations, the largest \$0.014; frozen IV and all LLY variants passed every check at the \$0.01 threshold. Counts were unchanged at 201 and 401 steps (Supplementary Table~\ref{tab:strike_audit_depth}). The failures therefore persisted under tree refinement; passing these limited checks did not establish an arbitrage-free model.

\FloatBarrier

\FloatBarrier

Tests on later dates identified shifts in IV bias, and an exploratory refit with earnings inputs reduced LLY's bias after its August report. Across 53 next-snapshot folds, median RMSE was 9.92 volatility points, with a range of 7.68--19.25 (Supplementary Table~\ref{tab:walk_forward_extended}). The separate July-cutoff fit rose from 10.79 points in training to 14.75 on the 23 August--September sessions and overpredicted IV by 6.88 points on average (Supplementary Table~\ref{tab:chronological_surface}). Its data screening and test period differed from those of the folds, so the two errors were not directly comparable. The held-out bias was 0.88 points for GS and 4.61 for LLY. Around LLY's August 5 earnings report, the surface changed from underpredicting IV by 17.90 points on August 4 to overpredicting it on every observed date from August 6 onward (Supplementary Table~\ref{tab:mechanism_checks}). In a subsequent exploratory refit, earnings inputs reduced LLY's average bias over August 6--September 4 from 6.17 to 3.85 points. The 38\% reduction fell short of the protocol's halving criterion, and substantial bias remained (Supplementary Table~\ref{tab:earnings_refit}). To keep these initial level errors out of the option comparison, every forecast variant started from observed IV and used a surface target rescaled to match it.

\begin{figure}[H]
\centering
\includegraphics[width=\textwidth]{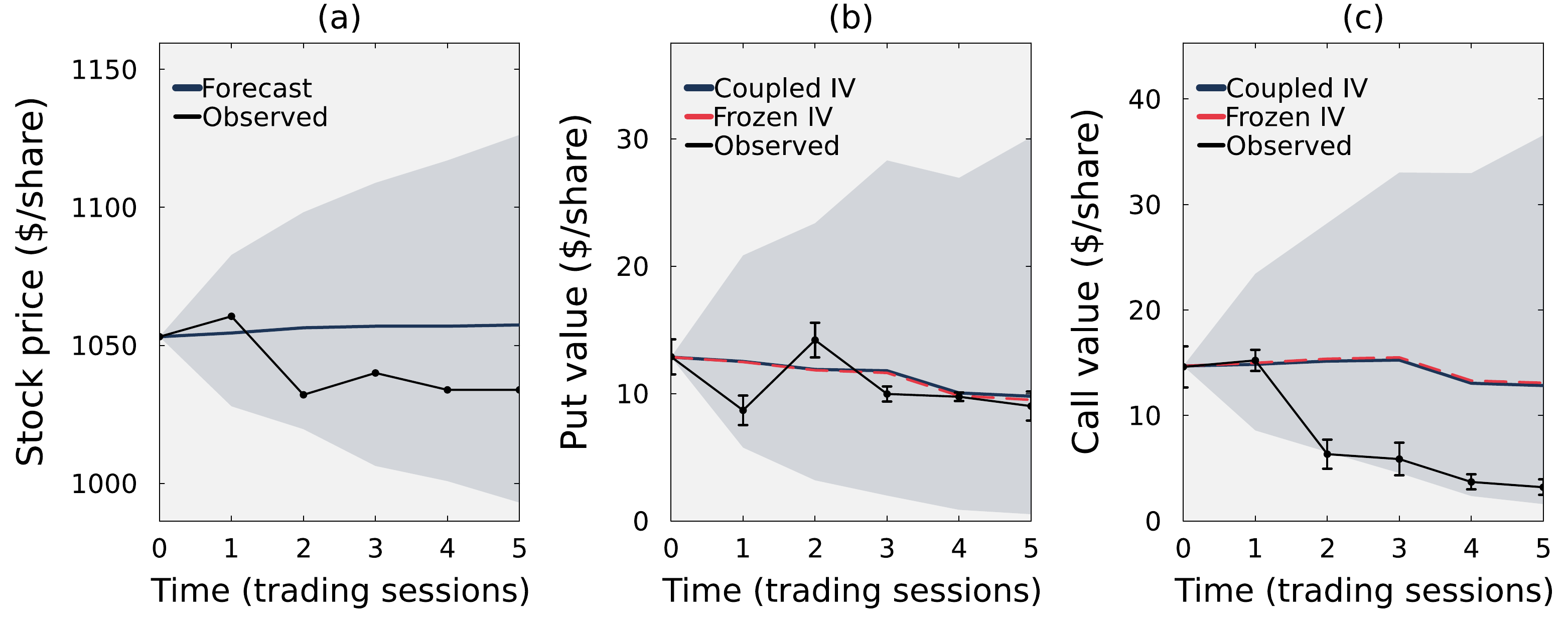}
\caption{Five-session forecasts of GS stock prices (a), put values (b), and call values (c), compared with observations from August 4 to August 11, 2026. The put strike is \$1{,}000, the call strike is \$1{,}105, and both options expire on August 21. Forecasts use the July-cutoff financials surface and information available at the August 4 origin. Navy curves show forecast means from 1{,}000 stock paths, with the full factor used for option values. Shading spans the central 90\% of simulated values. Red dashed curves show frozen-IV mean option forecasts on the same stock paths. Black points show observed stock closes and option midpoints, with vertical bars spanning Alpaca's indicative bid and ask for the options. Time is measured in trading sessions from the origin. August 4 was the earliest GS origin with both selected shorter-maturity options observed at the five-session endpoint.}\label{fig:chronological_forecast}
\end{figure}

In the five-session GS and LLY tests, the full IV update produced mean option-price forecasts with errors close to those under fixed IV (Table~\ref{tab:forecast_main}). The five-session comparison used the shorter-maturity cohort: the monthly cohort had no exact-contract outcomes at that horizon (Supplementary Table~\ref{tab:forecast_availability}). Seventeen origins per ticker supplied 34 GS and 30 LLY quotes. Compared with frozen IV, the full factor reduced mean-forecast MAE only from \$4.21 to \$4.19 per share for GS and from \$11.62 to \$11.59 for LLY. It improved on only six and nine of the seventeen origins, respectively. Median-forecast errors also showed no advantage, and one-session monthly forecasts gave similarly small differences (Supplementary Tables~\ref{tab:option_point_summary} and~\ref{tab:forecast_one_session}). The full factor's distribution score, CRPS, was close to frozen IV's, while its central 90\% intervals covered all GS outcomes but only 67.6\% of LLY outcomes. The example forecast likewise showed nearly coincident full-factor and frozen-IV means (Fig.~\ref{fig:chronological_forecast}). Over five sessions, average full-factor IV moved only 0.30 volatility points for GS and 0.18 for LLY, although individual paths moved by about 2.5 points in absolute terms (Supplementary Table~\ref{tab:mechanism_checks}). The slow update closed only about a quarter of the implied-variance gap to a fixed target, and shocks partly canceled across paths. These small mean IV changes were consistent with the similar option forecasts despite larger changes on individual paths.

\begin{table}[H]
\centering
\caption{Five-session option forecasts for the shorter-maturity cohort, selected with 10--21 calendar DTE at each origin. Scores average matched options within each ticker and origin before averaging origins. The MAE scores the forecast mean. Monetary quantities (MAE, CRPS, and central 90\% interval width) are in dollars per share, and coverage is date-weighted. The column \emph{Dates better} counts the origins on which a variant's date-level MAE was below that of frozen IV. Every IV variant uses the same stock paths, origin IV, and matched quotes.}\label{tab:forecast_main}
\resizebox{\textwidth}{!}{\begin{tabular}{llrrrrrrr}
\toprule
Ticker & IV variant & Dates & Quotes & MAE & CRPS & Coverage (\%) & Width & Dates better \\
\midrule
GS & Frozen IV & 17 & 34 & 4.21 & 2.02 & 100.0 & 25.64 & -- \\
GS & Direct surface & 17 & 34 & 4.86 & 2.16 & 67.6 & 24.01 & 2/17 \\
GS & Mean reversion & 17 & 34 & 4.26 & 2.02 & 100.0 & 25.62 & 1/17 \\
GS & Uncoupled factor & 17 & 34 & 4.24 & 2.03 & 100.0 & 25.69 & 3/17 \\
GS & Full factor & 17 & 34 & 4.19 & 2.01 & 100.0 & 25.65 & 6/17 \\
LLY & Frozen IV & 17 & 30 & 11.62 & 8.99 & 73.5 & 28.98 & -- \\
LLY & Direct surface & 17 & 30 & 11.62 & 9.05 & 55.9 & 26.68 & 7/17 \\
LLY & Mean reversion & 17 & 30 & 11.62 & 8.99 & 73.5 & 28.83 & 7/17 \\
LLY & Uncoupled factor & 17 & 30 & 11.61 & 8.97 & 73.5 & 28.83 & 9/17 \\
LLY & Full factor & 17 & 30 & 11.59 & 8.93 & 67.6 & 28.91 & 9/17 \\
\bottomrule\end{tabular}
}
\end{table}

Using observed stock paths greatly reduced option errors, showing the importance of stock forecasting to the joint forecast (Supplementary Table~\ref{tab:forecast_conditional}). On the thirteen origins per ticker with complete stock histories, the full-factor MAE fell from \$4.62 to \$1.23 for GS and from \$10.61 to \$1.95 for LLY. These were paired comparisons on the same 26 GS and 24 LLY quotes; the observed paths were future information and served only as a diagnostic. Repricing with both endpoint stock prices and reported IVs gave much smaller MAEs, \$0.14 and \$0.34, although that check used all seventeen matched origins rather than thirteen (Supplementary Table~\ref{tab:forecast_repricing}). Tree refinement changed sampled prices by at most \$0.0026 (Supplementary Table~\ref{tab:forecast_numerical_checks}). Together, these checks pointed to stock and IV prediction rather than tree accuracy as the main sources of error. Even with stock paths known, frozen IV had lower MAE than the full factor: \$1.07 versus \$1.23 for GS and \$1.74 versus \$1.95 for LLY. Random IV shocks did improve conditional distribution scores relative to frozen IV, but the uncoupled factor's CRPS of \$0.71 and \$1.35 was lower than the full factor's \$0.90 and \$1.53 (Supplementary Table~\ref{tab:forecast_conditional}). Coupling therefore did not improve these conditional distributions, and no IV update consistently improved point forecasts.

\FloatBarrier

Adaptive volatility improved LLY's stock forecast distributions relative to JumpHMM (Table~\ref{tab:stock_validation}). In the original option test, JumpHMM's mean-based stock errors were slightly higher than a Gaussian random walk's (Supplementary Table~\ref{tab:forecast_stock}). Reconstructing prices found no growth-to-price conversion error, and initializing the state from past returns had little effect on five-session median errors (Supplementary Table~\ref{tab:stock_initialization}). Historical validation selected an EWMA decay of $0.97$ and zero regression drift, making the direction-predicting model identical to the adaptive-volatility model (Supplementary Table~\ref{tab:stock_selection}). Neither it nor JumpHMM materially improved on an unchanged price in 2025 or the exploratory 2026 comparison; one-session MAE and RMSE were within 1\% of the benchmark for every method (Supplementary Tables~\ref{tab:stock_validation_five} and~\ref{tab:stock_validation_one}). For LLY in 2025, however, adaptive volatility improved on JumpHMM: it lowered five-session CRPS from \$29.60 to \$28.57 and raised central 90\% coverage from 69.9\% to 84.4\%, while widening intervals from \$100.40 to \$151.37. The 2026 LLY comparison changed in the same direction, with CRPS falling from \$35.60 to \$33.81 and coverage rising from 70.6\% to 94.1\%. For GS, adaptive volatility had slightly higher CRPS in 2025 and 20\% higher CRPS in 2026, \$15.58 versus \$13.03. Both models already covered every GS endpoint in 2026, so wider intervals added no coverage. Adaptive volatility tracked LLY's observed daily volatility more closely than JumpHMM's fixed spread but remained too high for GS in 2026 (Supplementary Table~\ref{tab:mechanism_checks}). It helped forecast uncertainty for one ticker without establishing a general advantage or improving direction forecasts.

\begin{table}[H]
\centering
\caption{Five-session stock forecasts in 2025 and August--September 2026. Each ticker has 245 origins in 2025 and 17 in 2026. Unlike Table~\ref{tab:forecast_main}, this table's MAE scores the forecast median. Monetary quantities (MAE, CRPS, and central 90\% interval width) are in dollars per share. Scores average dates within each of three 10{,}000-path simulation seeds and then average the seeds. Adaptive volatility denotes the Gaussian model with EWMA decay $0.97$ selected on historical data. Historical selection set the direction-predicting model's drift to zero, which made it identical to the adaptive-volatility model, so the table omits it. The unchanged-price benchmark has no uncertainty interval. The last column counts origins on which the adaptive-volatility model had lower date-level CRPS than JumpHMM.}\label{tab:stock_validation}
\resizebox{\textwidth}{!}{\begin{tabular}{lllrrrrr}
\toprule
Year & Stock & Model & Median MAE & CRPS & Coverage (\%) & Width & Dates with lower CRPS \\
\midrule
2025 & GS & JumpHMM & 22.31 & 16.22 & 84.4 & 84.71 & -- \\
2025 & GS & Unchanged & 22.61 & 22.61 & -- & -- & -- \\
2025 & GS & Adaptive volatility & 22.62 & 16.32 & 89.8 & 94.86 & 103/245 \\
2025 & LLY & JumpHMM & 39.10 & 29.60 & 69.9 & 100.40 & -- \\
2025 & LLY & Unchanged & 39.37 & 39.37 & -- & -- & -- \\
2025 & LLY & Adaptive volatility & 39.35 & 28.57 & 84.4 & 151.37 & 131/245 \\
2026 & GS & JumpHMM & 16.25 & 13.03 & 100.0 & 129.05 & -- \\
2026 & GS & Unchanged & 15.51 & 15.51 & -- & -- & -- \\
2026 & GS & Adaptive volatility & 15.48 & 15.58 & 100.0 & 176.18 & 2/17 \\
2026 & LLY & JumpHMM & 49.44 & 35.60 & 70.6 & 147.26 & -- \\
2026 & LLY & Unchanged & 49.37 & 49.37 & -- & -- & -- \\
2026 & LLY & Adaptive volatility & 49.31 & 33.81 & 94.1 & 191.38 & 10/17 \\
\bottomrule
\end{tabular}
}
\end{table}

Widening the JumpHMM return cutoff from $b=10$ to $b=20$ left these conclusions unchanged (Supplementary Tables~\ref{tab:tail_sensitivity} and~\ref{tab:tail_forecasts}). In the illustrations, mean short-call P\&L at expiration changed by less than \$0.02 per share and its 5\% expected shortfall by less than \$0.07. At the ablation's ten-day endpoint, the corresponding full-factor changes were below \$0.004 and \$0.02. Five-session full-factor option MAE changed by less than \$0.007, and JumpHMM stock CRPS by less than \$0.002. The primary cutoff removed only about 0.017\% of the original emission probability (Supplementary Table~\ref{tab:emission_bounds}). Because so few draws approached either bound, this stability did not establish robustness to rare returns beyond them.

\FloatBarrier

\section{Discussion}\label{sec:discussion}

Separating the IV update from the stock model allowed controlled comparisons of option values under different IV assumptions. We compared five IV variants for short positions in four out-of-the-money one-month options: a put and a call on each of Goldman Sachs (GS) and Eli Lilly (LLY), with entry premiums set to their April 28, 2026 quote midpoints. In this April 28 comparison, we generated stock paths with the jump hidden Markov model JumpHMM and repriced the options along the same paths under each variant. Frozen IV held each option's IV at its starting value. The direct surface set IV each day to the prediction of a neural IV surface fitted to that ticker, and we call the square of this prediction the target. In the other three variants, each option carried its own simulated implied variance, its IV factor, which moved partway toward the target each day. The mean-reversion variant made only this move. The uncoupled factor added random shocks unrelated to the stock, and the full factor, the complete update we propose, added shocks correlated with stock returns. Because the variants shared the stock paths, the options, and the pricing tree, any difference in option values came from the IV variant alone. All five started from the target, so their values on the entry date agreed with one another, although not with the quoted midpoints. We did not model early assignment, so payoffs at expiration depended only on the final stock price and also agreed. We valued the positions after ten trading days and measured each variant against the direct surface, which followed the target with no lag and no shocks. Differences from it therefore showed the effects of lag and shocks. The surfaces were fitted on quotes through May 11, after the entry date, so this comparison shows how the variants behave rather than forecasting prices.

The IV dynamics changed prices and tail losses before expiration even when average profit and loss (P\&L) changed little. The mean-reversion variant left prices nearly as far from those of the direct surface as frozen IV did (Supplementary Table~\ref{tab:dynamic_ablation}). The assumed mean-reversion rate closed only about 6\% of the gap to the target each trading day, while the target moved with the stock price and the time to expiration. Random shocks then moved prices on individual paths much more than they changed average P\&L, because their effects largely canceled across paths. Correlating the shocks with stock returns changed the tail of the P\&L. The full factor linked shocks to stock returns with a parameter set to $-0.6$ by assumption. As a result, IV tended to rise on days the stock fell, consistent with the leverage effect~\cite{cont2001}. On falling-stock paths, higher IV raised put values and deepened the losses of short puts. On rising-stock paths, lower IV reduced call values and softened the losses of short calls (Fig.~\ref{fig:dynamic_ablation}). Compared with the uncoupled factor at the fixed ten-day horizon, the average loss in the worst 5\% of outcomes increased for both short puts and decreased for both short calls. Each change was at least seven times its simulation standard error (Supplementary Table~\ref{tab:dynamic_ablation}). These comparisons showed how IV assumptions affected simulated losses at a fixed horizon, even when mean P\&L changed little and expiration payoffs agreed. They established sensitivity to the chosen dynamics; they did not test an exit rule or establish that those dynamics matched the market.

More flexible IV surfaces improved the fit to calibration quotes, and adding earnings information reduced some errors on withheld dates. The per-ticker surfaces were the closest fits to the calibration quotes (Table~\ref{tab:three_way}), yet they overpredicted near-the-money IV for Microsoft (MSFT) and LLY on May 11, the last fitting date (Fig.~\ref{fig:per_ticker_panels}). Option prices have long shown IV rising before an anticipated announcement and falling after it~\cite{patell1979,dubinsky2019}. None of the surfaces used by the simulator had an earnings input, and each held a ticker's IV level constant across its fitting dates. Ten of the fifteen calibration snapshots preceded both companies' April reports, so the fitted level could remain above the IV observed afterward. The sector surfaces fitted through July 31 likewise underpredicted LLY's IV before its August 5 report and overpredicted it on every observed day of the following month (Supplementary Table~\ref{tab:mechanism_checks}). Adding earnings inputs reduced errors in the April holdout and reduced LLY's average overprediction after the August report by 38\% in the exploratory refit, falling short of the protocol's halving criterion (Table~\ref{tab:earnings_holdout}; Supplementary Table~\ref{tab:earnings_refit}). Adding earnings inputs reduced LLY's bias, although substantial error remained. Accurate individual prices would also be insufficient for strategies holding several options, whose decisions depend on relative prices. Under the direct surface, simulated GS prices failed 178 of 1{,}818 convexity checks by more than \$0.01 per share, implying negative-priced butterfly spreads (Supplementary Table~\ref{tab:strike_audit}). We did not constrain the surfaces to be free of arbitrage, and we priced each option separately. This separation made the effects of IV assumptions easy to isolate but left consistency across strikes unenforced. The static no-arbitrage constraints used in methods that simulate IV surfaces offer a starting point~\cite{cont2023}. Those constraints are stated for European prices, whereas the simulator prices American options, so its prices would still need checks like ours.

The forecast diagnostics identified stock prediction as a major source of option-price error, while the stock benchmarks showed that adapting volatility improved LLY's forecast distributions. In the five-day forecasts, the full factor was only slightly more accurate than frozen IV when scored by the mean absolute error of the average over paths, and its small advantage was not consistent across dates (Table~\ref{tab:forecast_main}). Scored by the median over paths, it was no more accurate than frozen IV (Supplementary Table~\ref{tab:option_point_summary}). All variants started from observed IV, with each target rescaled to match that starting value, keeping initial surface-level errors out of the comparison. The slow update closed only about a quarter of a gap to a fixed target within five trading days, and random shocks partly canceled across paths, so average IV stayed near the value that frozen IV held fixed (Supplementary Table~\ref{tab:mechanism_checks}). With a fixed surface, rate, shock size, and correlation, the update could not anticipate shifts in the observed surface's level, skew, and curvature~\cite{cont2002}. Stock prediction was another major source of error: repeating forecasts along the stock paths that actually occurred cut the full factor's option errors 3.8-fold for GS and 5.4-fold for LLY on the thirteen dates per ticker with complete histories (Supplementary Table~\ref{tab:forecast_conditional}). This was a diagnostic using future information, and frozen IV still had lower errors on those paths. Neither JumpHMM nor the adaptive-volatility stock model materially improved point forecasts over assuming no price change, while historical tuning set the direction-predicting model's drift to zero (Table~\ref{tab:stock_validation}; Supplementary Table~\ref{tab:stock_selection}). The adaptive-volatility model improved LLY's forecast distributions in both 2025 and 2026, but worsened GS's. The LLY improvement was consistent with evidence that volatility is partly predictable from recent returns~\cite{cont2001,andersen1998}. Its wider ranges tracked LLY's higher volatility, whereas JumpHMM and the adaptive-volatility model already covered every GS outcome in 2026, so wider ranges added no coverage~\cite{gneiting2007calibration}. Improving forecast uncertainty was therefore distinct from predicting direction, and neither the IV update nor the stock alternatives established a general forecasting advantage.

The simulator provided a reproducible way to compare IV and stock assumptions, with conclusions bounded by the available data and modeling choices. All option quotes came from Alpaca's free indicative feed, which is derived from exchange quotes. Indicative quotes are not firm offers, so our price errors measured agreement with Alpaca's quotes, not with prices at which trades could execute. Stock and option prices were also not recorded at exactly the same moment, which may affect every IV and price error. The observation windows were short, and five-day option forecasts were scored on only 17 sessions. Overlapping forecast periods and shared market moves further reduced the independent information in these tests. Missing exact-contract outcomes also limited five-day forecasts to shorter-maturity options; complete daily histories would allow the test to be repeated on one-month options. Apart from the earnings refit, each network tested on later dates was trained with a single random seed. We proposed the alternative stock models after seeing JumpHMM's August and September forecast misses, so their 2026 comparison was not a blind test. The earnings explanation rested on a few reports and two tests that left substantial error. Widening the cutoff on JumpHMM's daily returns changed the reported scores little, but did not establish robustness to the largest real stock moves. The IV update's rate, shock size, and correlation were assumed rather than estimated, and its target level did not depend on the stock model's hidden state. Quotes spanning calm and turbulent markets would allow those state-dependent levels to be estimated and tested. Together with prices consistent across strikes and better forecasts of changes in IV and stock uncertainty, these data would support a stronger assessment of the simulator. The present evidence supports comparisons of modeling assumptions, not trading decisions, and its forecast conclusions apply only to the options and dates tested here.

\section{Conclusion}\label{sec:conclusion}

In this study, we built a simulator that values American options along stock paths from a separately chosen stock model, here the jump hidden Markov model JumpHMM. On every simulated date, the simulator moves each option's implied variance partway toward a target predicted by a neural surface fitted to market quotes. This update borrows the mean-reverting form of the Heston model's variance equation and adds random shocks. A binomial tree then converts the resulting IV into the option's price. More flexible surfaces, fitted separately to smaller groups of tickers, matched the fitting data more closely, but their accuracy on later dates was uneven. Adding earnings information reduced some held-out errors, including LLY's bias after its August report. Repricing identical stock paths under different IV updates changed option values and hypothetical short-position losses at a fixed horizon of ten trading days, while payoffs at expiration stayed the same. At that horizon, linking the random shocks to stock returns raised the worst simulated losses of short puts and lowered those of short calls. In forecasts of later option prices, the IV update with random shocks linked to stock returns showed no consistent advantage over holding IV fixed. Repeating the forecasts with the stock paths that actually occurred reduced option errors, which pointed to the stock forecast as the main source of error. In stock forecasts for 2025 and for August and September 2026, JumpHMM, an adaptive-volatility model, and a direction-predicting model all predicted prices about as accurately as assuming no change. The adaptive-volatility model improved the predicted range of stock prices for LLY in both periods but not for GS. Tuning on 2019--2024 data found no benefit from predicting direction with recent stock and market returns, so the direction-predicting model, which adds a drift to the adaptive-volatility model, gave the same forecasts. The simulator therefore supports reproducible comparisons of IV and stock assumptions, but it did not forecast better than simpler alternatives. Saved observations, fitted models, simulation outputs, and scoring code make every comparison reviewable. Using these paths to assess trading or hedging rules would require further validation against observed option-price variation and prices consistent across strikes.

\section*{Data and Code Availability}\label{sec:data_availability}

All data, code, and saved results for this study are available at \url{https://github.com/varnerlab/heston_implied_volatility_model}. The repository contains the daily option-chain snapshots from April 14 through September 4, 2026, the daily stock prices, the earnings calendar, and the pretrained JumpHMM model. It also contains the scripts that fit the IV surfaces, run the simulations and tests, and summarize their results, together with the saved outputs and logs. The repository's \texttt{README.md} gives the command for each experiment, and \texttt{SETUP.md} lists the setup steps. The option quotes come from Alpaca's free indicative feed, a modified version of the exchange feed, so they may differ from exchange quotes~\cite{alpaca_option_feed}. The code is released under the MIT License.

\bibliographystyle{abbrvnat}
\setlength{\bibsep}{3pt}
\bibliography{references}

\clearpage

\appendix
\renewcommand{\thefigure}{S\arabic{figure}}
\renewcommand{\thetable}{S\arabic{table}}
\renewcommand{\thealgorithm}{S\arabic{algorithm}}
\setcounter{figure}{0}
\setcounter{table}{0}
\setcounter{algorithm}{0}
\section{Interpretation of the IV Factor}\label{sec:supp_sigma_decomp}

A continuous-time square-root process motivated the Euler update in~\eqref{eq:euler}. The continuous-time process is given by:
\begin{equation}\label{eq:level_sde}
    dv_{a,t}=\kappa(\bar v_{a,t}-v_{a,t})\,dt
    +\sigma_v\sqrt{v_{a,t}}\,dW_{a,t},
\end{equation}
where contract $a$ indexes ticker, strike, and expiration, and the target $\bar v_{a,t}$ from~\eqref{eq:theta} changes with current moneyness and calendar DTE. With the target held constant at $\bar v_a$, the stationary mean, when it exists, equals that target. The simulation instead used a moving target, discrete Euler steps, a variance floor, and non-normal stock-linked shocks. Mean reversion motivated the update, but the fixed-target stationary distribution does not describe the simulated factors.

\section{Calibration Details}\label{sec:supp_sector_nn}

The \href{https://github.com/varnerlab/heston_implied_volatility_model}{source repository} provides the experiment drivers and reproduction instructions. Its data manifest records each file's SHA-256 hash and filtered row count. The saved ablation outputs include the pilot constants, path-level prices, numerical checks, and copies of the source code, so each reported comparison can be traced to its inputs.

Each calibration snapshot had a label, a capture timestamp, and an underlying-session date, and these did not always coincide (Table~\ref{tab:corpus_dates}). The distinction matters because the holdout tests split dates in different ways. All dates are in 2026. For example, the snapshot for the May 11 underlying session was captured on May 12. The first snapshot contained 23 tickers, and each later snapshot contained 31.
\begin{table}[H]
\centering
\caption{Labels, capture dates, and underlying-session dates of the calibration snapshots. The April earnings holdout split dates by snapshot label. The May 11 single-date holdout used the underlying-session date. The first snapshot is labeled April 14, but its underlying close was from April 13. The Rows column counts filtered observations from each snapshot.}\label{tab:corpus_dates}
\begin{tabular}{lllr}
\toprule
Snapshot label & Capture date & Underlying session & Rows \\
\midrule
04-14-2026 & 2026-04-14 & 2026-04-13 & 18,250 \\
04-15-2026 & 2026-04-15 & 2026-04-15 & 17,711 \\
04-16-2026 & 2026-04-16 & 2026-04-16 & 18,344 \\
04-17-2026 & 2026-04-17 & 2026-04-17 & 17,281 \\
04-21-2026 & 2026-04-21 & 2026-04-21 & 16,760 \\
04-22-2026 & 2026-04-22 & 2026-04-22 & 15,551 \\
04-23-2026 & 2026-04-24 & 2026-04-23 & 11,482 \\
04-24-2026 & 2026-04-24 & 2026-04-24 & 15,351 \\
04-27-2026 & 2026-04-27 & 2026-04-27 & 11,379 \\
04-28-2026 & 2026-04-28 & 2026-04-28 & 16,776 \\
04-29-2026 & 2026-04-30 & 2026-04-30 & 16,811 \\
05-01-2026 & 2026-05-04 & 2026-05-01 & 14,825 \\
05-06-2026 & 2026-05-07 & 2026-05-06 & 17,293 \\
05-08-2026 & 2026-05-08 & 2026-05-08 & 12,293 \\
05-11-2026 & 2026-05-12 & 2026-05-11 & 14,442 \\
\bottomrule
\end{tabular}

\end{table}
\FloatBarrier

\begin{table}[H]
\centering
\caption{Sector assignment of the 31 tickers across the fifteen snapshots. Counts are the filtered observations used in fitting.}\label{tab:sector_map}
\resizebox{\textwidth}{!}{%
\begin{tabular}{lrl}
\toprule
Sector & Observations & Tickers \\
\midrule
ETF        & 102{,}738 & IWM, QQQ, SPY \\
Technology &  68{,}334 & AAPL, AMD, AVGO, GOOG, INTC, META, MSFT, MU, NVDA, QCOM \\
Healthcare &  26{,}387 & ABBV, AMGN, BMY, JNJ, LLY, MRNA, PFE, UNH \\
Financials &  17{,}938 & BAC, GS, JPM, WFC \\
Retail     &  11{,}320 & TGT, UPS, WMT \\
Energy     &   7{,}832 & CVX, OXY, XOM \\
\bottomrule
\end{tabular}}
\end{table}

\begin{table}[H]
\centering
\caption{In-sample IV RMSE of the parametric, global neural, and sector neural surfaces by sector. Errors are in volatility percentage points. The sector network improved on the global neural network in every group, but technology and healthcare kept the largest errors.}\label{tab:per_sector}
\begin{tabular}{lrrrr}
\toprule
Sector & Parametric & Global NN & Sector NN & Observations \\
\midrule
ETF        & 10.54 &  8.69 &  6.10 & 102{,}738 \\
Financials &  9.05 &  7.79 &  6.70 &  17{,}938 \\
Energy     &  9.36 &  8.24 &  7.88 &   7{,}832 \\
Retail     & 11.73 & 10.23 &  9.50 &  11{,}320 \\
Healthcare & 13.49 & 12.71 & 12.53 &  26{,}387 \\
Technology & 15.58 & 15.31 & 14.47 &  68{,}334 \\
\bottomrule
\end{tabular}
\end{table}

The per-ticker comparison reflected network size as well as sharing across stocks (Table~\ref{tab:per_ticker_comparison}). Sector networks improved on the global network in all six groups, but technology and healthcare retained the largest errors and weakest individual fits (Tables~\ref{tab:per_sector} and~\ref{tab:worst_tickers}). Of the 29 tickers with their own network, 21 improved on their sector fit. The eight exceptions were financials, energy, or retail stocks with fewer than 5{,}000 observations, which received 105-parameter networks rather than the sector networks' 337 parameters. Refitting all 18 small networks at the larger size with the same data, seed, and schedule made every refit outperform its sector fit. Retaining final rather than lowest-loss weights did not explain the losses: those errors differed by less than 0.001 volatility points. These diagnostics supported network size as the explanation for the eight exceptions, but did not establish better out-of-sample performance. We retained the original size rule, chosen to limit overfitting. The per-ticker surfaces used in the simulations, including GS and LLY, already had the larger architecture.

\begin{table}[H]
\centering
\caption{In-sample IV RMSE of the sector and per-ticker networks for the 29 qualified tickers. Rows are sorted by the reduction from the sector network to the ticker's own network. A positive reduction means the ticker's own network fitted better than its sector network. Units gives the width of each hidden layer of the ticker's own network, 16 for tickers with at least 5{,}000 observations and 8 otherwise. The 16-unit refit column repeats each 8-unit fit with 16-unit layers and the same data, seed, and training schedule. These refits served only as a diagnostic and entered no other result. The final rows count the tickers that improved and worsened and the refits that fitted better than their sector network. Two tickers, BMY and PFE, kept their sector surfaces. Combining per-ticker networks with these sector fallbacks reduced pooled RMSE from 10.24 to 9.73 volatility percentage points (Table~\ref{tab:three_way}).}\label{tab:per_ticker_comparison}
\begin{tabular}{lrrrrrr}
\toprule
Ticker & Observations & Units & Sector NN & Per-ticker NN & Reduction & 16-unit refit \\
\midrule
JNJ & 2{,}190 & 8 & 11.14 & 7.17 & 3.97 & 6.80 \\
MRNA & 2{,}851 & 8 & 17.49 & 13.74 & 3.76 & 13.46 \\
AAPL & 6{,}594 & 16 & 14.30 & 12.59 & 1.70 & -- \\
INTC & 3{,}417 & 8 & 17.78 & 16.29 & 1.49 & 15.99 \\
AMGN & 3{,}383 & 8 & 13.36 & 12.14 & 1.21 & 11.70 \\
MU & 6{,}619 & 16 & 18.45 & 17.26 & 1.19 & -- \\
NVDA & 5{,}805 & 16 & 10.14 & 9.15 & 0.99 & -- \\
AVGO & 9{,}470 & 16 & 9.94 & 9.28 & 0.67 & -- \\
UNH & 3{,}760 & 8 & 10.39 & 9.77 & 0.63 & 9.28 \\
AMD & 4{,}566 & 8 & 17.74 & 17.13 & 0.61 & 17.03 \\
MSFT & 10{,}129 & 16 & 14.33 & 13.79 & 0.54 & -- \\
GOOG & 4{,}433 & 8 & 9.57 & 9.05 & 0.53 & 8.72 \\
QCOM & 3{,}375 & 8 & 21.46 & 20.99 & 0.47 & 20.32 \\
GS & 8{,}772 & 16 & 5.40 & 4.98 & 0.42 & -- \\
SPY & 42{,}224 & 16 & 6.02 & 5.72 & 0.30 & -- \\
IWM & 21{,}368 & 16 & 6.11 & 5.82 & 0.29 & -- \\
LLY & 8{,}531 & 16 & 11.35 & 11.06 & 0.29 & -- \\
META & 13{,}926 & 16 & 13.72 & 13.45 & 0.26 & -- \\
ABBV & 2{,}525 & 8 & 9.13 & 8.95 & 0.18 & 8.63 \\
QQQ & 39{,}146 & 16 & 6.17 & 6.08 & 0.09 & -- \\
TGT & 3{,}957 & 8 & 9.54 & 9.48 & 0.06 & 9.11 \\
BAC & 2{,}399 & 8 & 9.47 & 9.53 & $-0.06$ & 8.31 \\
XOM & 2{,}820 & 8 & 7.31 & 7.41 & $-0.10$ & 7.21 \\
CVX & 2{,}453 & 8 & 7.70 & 7.81 & $-0.11$ & 7.53 \\
OXY & 2{,}559 & 8 & 8.60 & 8.81 & $-0.20$ & 8.34 \\
WFC & 3{,}047 & 8 & 7.94 & 8.17 & $-0.24$ & 7.63 \\
UPS & 3{,}526 & 8 & 11.44 & 11.68 & $-0.25$ & 11.09 \\
WMT & 3{,}837 & 8 & 7.23 & 7.52 & $-0.29$ & 6.70 \\
JPM & 3{,}720 & 8 & 6.18 & 6.65 & $-0.47$ & 5.97 \\
\midrule
\multicolumn{6}{l}{Tickers with lower per-ticker RMSE} & 21 \\
\multicolumn{6}{l}{Tickers with higher per-ticker RMSE} & 8 \\
\multicolumn{6}{l}{Refitted tickers with lower RMSE than their sector network} & 18 of 18 \\
\bottomrule
\end{tabular}

\end{table}

\begin{table}[H]
\centering
\caption{Observed and predicted mean IV near the money in the May 11 smile panels (Fig.~\ref{fig:per_ticker_panels}). Each row averages the plotted calls and puts with $0.98\le K/S\le1.02$ at the indicated calendar days to expiry (DTE). Predictions use the saved fifteen-date fit, evaluated at each observed contract. Observed and predicted IV are percentages. Bias is the mean per-ticker predicted IV minus observed IV, in percentage points. The NVDA subset contains only two observations. According to the public earnings calendar, GS, MSFT, and LLY reported on April 13, April 29, and April 30, 2026, respectively, all within the calibration window.}\label{tab:smile_date_bias}
\begin{tabular}{lrrrrr}
\toprule
Ticker & DTE & $N$ & Observed IV & Predicted IV & Bias \\
\midrule
SPY & 7 & 52 & 13.57 & 13.75 & +0.18 \\
NVDA & 7 & 2 & 38.65 & 34.02 & -4.62 \\
MSFT & 9 & 14 & 28.47 & 37.92 & +9.45 \\
LLY & 11 & 20 & 38.59 & 46.49 & +7.90 \\
GS & 11 & 24 & 32.06 & 31.74 & -0.32 \\
AVGO & 9 & 14 & 49.08 & 43.72 & -5.37 \\
\bottomrule\end{tabular}

\end{table}

\begin{table}[H]
\centering
\caption{Median half-spreads and American-option pricing errors on May 11 for the contracts plotted in Fig.~\ref{fig:price_error}. Dollar values are medians across the plotted contracts. The per-ticker error is the absolute difference between the tree price using fitted IV and the market mid. Inside spread is the share of contracts whose absolute per-ticker error was at most their half-spread. The market-IV reference passed each contract's reported IV to the same 200-step CRR tree.}\label{tab:price_error_summary}
\resizebox{\textwidth}{!}{%
\begin{tabular}{lrrrrrr}
\toprule
Ticker & DTE & $N$ & Median half-spread & Median per-ticker error & Inside spread & Median market-IV error \\
\midrule
SPY & 7 & 176 & \$0.03 & \$0.12 & 33\% & \$0.01 \\
NVDA & 7 & 34 & \$0.04 & \$0.13 & 35\% & \$0.12 \\
MSFT & 9 & 91 & \$0.36 & \$1.48 & 20\% & \$0.37 \\
LLY & 11 & 156 & \$2.98 & \$4.01 & 30\% & \$0.67 \\
GS & 11 & 145 & \$1.56 & \$0.46 & 88\% & \$0.12 \\
AVGO & 9 & 116 & \$0.81 & \$0.57 & 56\% & \$0.06 \\
\bottomrule
\end{tabular}
}
\end{table}

\begin{table}[H]
\centering
\caption{Five best and five worst tickers by in-sample IV RMSE under the sector networks. Errors are in volatility percentage points.}\label{tab:worst_tickers}
\begin{tabular}{llrrl}
\toprule
Rank & Ticker & Observations & RMSE & Sector \\
\midrule
\multicolumn{5}{l}{\emph{Best five}} \\
 1 & GS   &  8{,}772 & 5.40 & Financials \\
 2 & SPY  & 42{,}224 & 6.02 & ETF \\
 3 & IWM  & 21{,}368 & 6.11 & ETF \\
 4 & QQQ  & 39{,}146 & 6.17 & ETF \\
 5 & JPM  &  3{,}720 & 6.18 & Financials \\
\midrule
\multicolumn{5}{l}{\emph{Worst five}} \\
27 & PFE  &  1{,}300 & 17.66 & Healthcare \\
28 & AMD  &  4{,}566 & 17.74 & Technology \\
29 & INTC &  3{,}417 & 17.78 & Technology \\
30 & MU   &  6{,}619 & 18.45 & Technology \\
31 & QCOM &  3{,}375 & 21.46 & Technology \\
\bottomrule
\end{tabular}
\end{table}

\FloatBarrier
\subsection{Single-date holdout}\label{sec:supp_loo}

To test transfer to a nearby date, we held out May 11, 2026, and retrained the sector surfaces on the other fourteen dates, standardizing inputs with the training rows alone. Test error exceeded training error only slightly when pooled, but by more in two sectors (Table~\ref{tab:loo_sector_nn}). The pooled gap was 0.68 volatility percentage points, whereas energy and retail had larger gaps. The small pooled gap therefore hid poorer transfer in those sectors. The test used one held-out date and one training seed.

\begin{table}[H]
\centering
\caption{Training and test RMSE of the sector networks with May 11, 2026, held out. The networks were fitted on the other fourteen dates and tested on the 14{,}442 observations from May 11. Values and test-minus-train gaps are in volatility percentage points.}\label{tab:loo_sector_nn}
\begin{tabular}{lrrr}
\toprule
Sector & Train & Test & Gap \\
\midrule
ETF        &  6.12 &  5.87 & $-0.25$ \\
Financials &  6.72 &  6.32 & $-0.40$ \\
Healthcare & 12.58 & 12.00 & $-0.58$ \\
Technology & 14.45 & 15.04 & $+0.58$ \\
Energy     &  7.74 &  9.97 & $+2.23$ \\
Retail     &  9.21 & 13.17 & $+3.96$ \\
\midrule
Pooled     & 10.21 & 10.89 & $+0.68$ \\
\bottomrule
\end{tabular}
\end{table}

\FloatBarrier
\section{Earnings-Holdout Ticker Results}\label{sec:supp_earnings_diag}

We broke the April earnings-holdout result down by technology ticker and found that the earnings inputs lowered test error for some tickers and raised it for others (Table~\ref{tab:tech_pivot}). The two-input and four-input models used the same training and test observations. Several tickers with a peer earnings date near the holdout had lower error with the earnings inputs, but AVGO, NVDA, and INTC had higher errors. Because the ticker and peer distances entered together, the comparison could not separate their contributions. Calendar distance also carries no information about the sign or size of an earnings surprise.

\begin{table}[H]
\centering
\caption{Test RMSE of the technology tickers in the April earnings holdout under the pooled two-input sector model (A) and the earnings-aware model (C). The columns $e$ and $e^{\mathrm{peer}}$ give test-row medians of the signed ticker and absolute nearest-peer earnings distances, capped at 30 days in magnitude. Differences are C minus A, in volatility percentage points.}\label{tab:tech_pivot}
\begin{tabular}{lrrrrr}
\toprule
Ticker & $e$ & $e^{\mathrm{peer}}$ & A: pooled & C: event inputs & Difference \\
\midrule
QCOM & $+5$  & $1$ & 41.66 & 37.59 & $-4.07$ \\
AMD  & $+11$ & $1$ & 40.07 & 36.50 & $-3.57$ \\
META & $+5$  & $1$ & 14.32 & 10.73 & $-3.59$ \\
MSFT & $+5$  & $1$ & 14.00 & 10.73 & $-3.28$ \\
GOOG & $+5$  & $1$ &  9.87 &  7.90 & $-1.97$ \\
MU   & $-30$ & $1$ & 16.95 & 15.59 & $-1.36$ \\
AAPL & $+6$  & $1$ & 10.42 & 10.34 & $-0.08$ \\
AVGO & $+30$ & $1$ & 12.58 & 12.83 & $+0.24$ \\
NVDA & $+26$ & $1$ &  9.43 & 10.47 & $+1.05$ \\
INTC & $-1$  & $5$ & 33.17 & 35.94 & $+2.77$ \\
\bottomrule
\end{tabular}
\end{table}

\FloatBarrier
\section{Scenario Statistics and Greeks}\label{sec:supp_short_premium}

The GS and LLY illustrations shared one simulation design (Table~\ref{tab:scenario_settings}). For GS, the seller usually kept the full premium but faced large losses in the lower P\&L tail (Table~\ref{tab:gs_scenario}). The premium-retained interval is a two-sided 95\% Wilson interval for the fraction of paths that expired out of the money. It measures only the Monte Carlo sampling error in that fraction.

\begin{table}[H]
\centering
\caption{Simulation settings shared by the GS and LLY illustrations.}\label{tab:scenario_settings}
\begin{tabular}{lr}
\toprule
Setting & Value \\
\midrule
Simulation paths & 1{,}000 \\
Random seed & 20260429 \\
Horizon & April 28--May 29, 2026 \\
Calendar days & 31 \\
Trading transitions & 22 \\
Annual mean-growth prior & 10\% \\
Mean-reversion rate $\kappa$ & 15 \\
Factor volatility $\sigma_v$ & 0.5 \\
Return--factor coupling $\rho$ & $-0.6$ \\
Risk-free rate $r$ & 4.25\% \\
Dividend yield $q$ & 0 \\
Leisen--Reimer depth & 201 \\
\bottomrule
\end{tabular}
\end{table}

\begin{table}[H]
\centering
\caption{Entry prices and short-position outcomes for the GS put and call over 1{,}000 simulated paths from April 28 to May 29, 2026. The horizon had 31 calendar days and 22 trading-day steps. Option values and P\&L are in dollars per share.}
\label{tab:gs_scenario}
\begin{tabular}{lrr}\toprule
Statistic & Short put & Short call \\\midrule
Initial spot & \multicolumn{2}{c}{926.38} \\
Strike & 890.00 & 970.00 \\
Market delta & -0.295 & +0.328 \\
Market IV & 31.2\% & 28.9\% \\
Market mid & 16.51 & 16.09 \\
Model entry value & 17.26 & 16.07 \\
Model value minus market mid & 0.75 & -0.02 \\
Mean P\&L & 4.30 & -2.19 \\
Monte Carlo SE of mean & 0.89 & 1.23 \\
Median P\&L & 16.51 & 16.09 \\
Standard deviation & 28.01 & 38.87 \\
5\% quantile & -58.96 & -86.18 \\
5\% expected shortfall & -92.15 & -130.77 \\
Premium retained & 74.0\% & 67.7\% \\
Wilson 95\% interval & 71.2--76.6\% & 64.7--70.5\% \\
Delta-heuristic premium retained & 70.5\% & 67.2\% \\
Simulated minus heuristic & +3.5 pp & +0.5 pp \\
\bottomrule\end{tabular}

\end{table}

\begin{figure}[H]
\centering
\includegraphics[width=\textwidth]{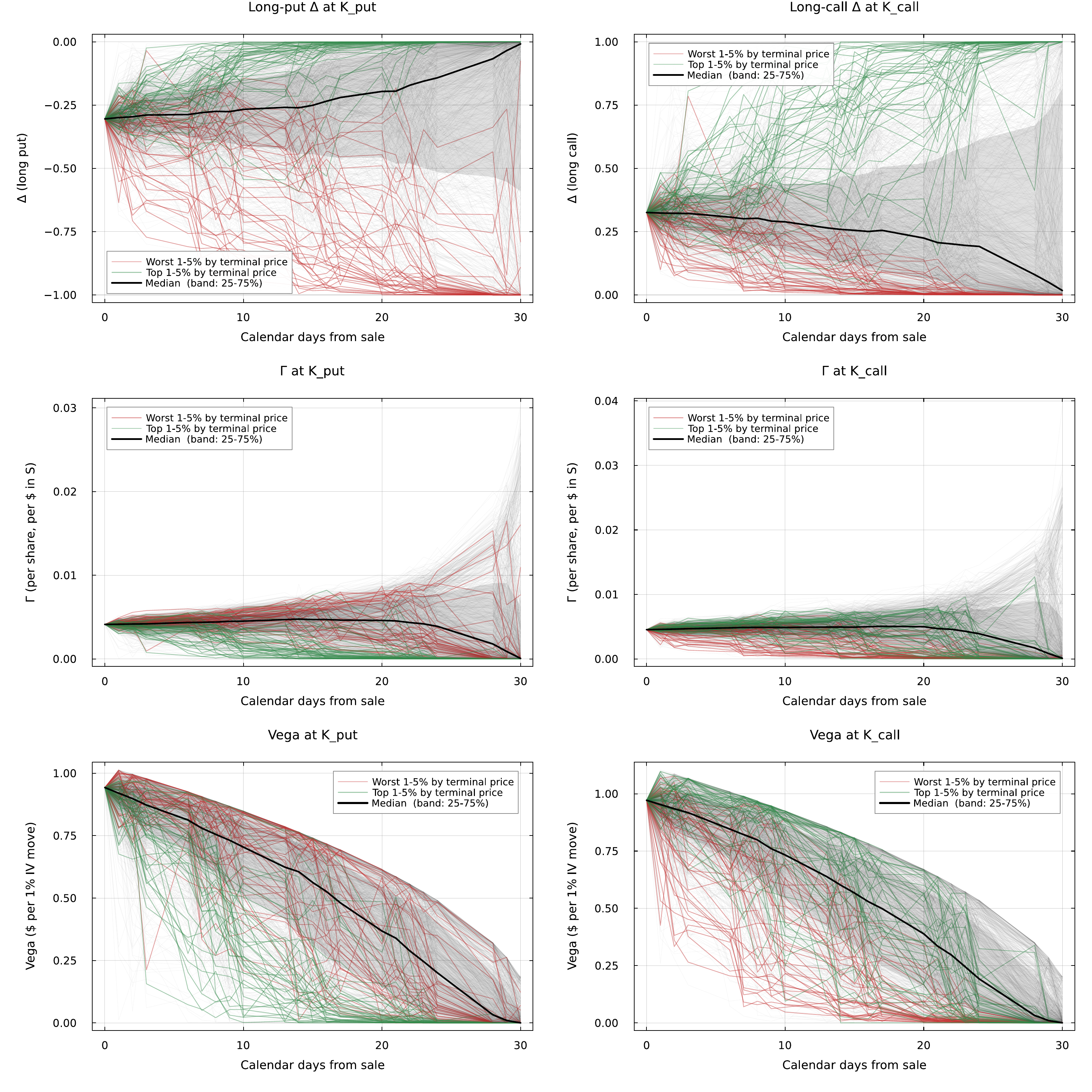}
\caption{Simulated evolution of GS long-option delta, gamma, and Vega along the stock paths. Sensitivities come from the 201-step Leisen--Reimer tree, with stock-price bumps of $1.5\%$ and IV bumps of 0.005. Vega is in dollars per volatility percentage point.}
\label{fig:gs_greeks}
\end{figure}

\FloatBarrier
We calculated Greeks along each path by central finite differences on the same Leisen--Reimer tree that supplied the base price (Algorithm~\ref{alg:greeks}). For GS, delta approached $\pm1$ on paths heading toward exercise, gamma concentrated near the strike, and Vega declined toward zero (Fig.~\ref{fig:gs_greeks}). The figures use the long-option sign convention, so a short position's Greeks have the opposite sign.
\begin{algorithm}[H]
\caption{Greeks by central finite differences along each path}\label{alg:greeks}
\begin{algorithmic}[1]
\Require State $(S,\sigma,\tau)$ with calendar DTE $\tau$; contract $(K,r)$; LR tree steps $N_{\mathrm{LR}}$; relative spot bump $h_S$; absolute IV bump $h_\sigma$.
\Ensure $\Delta$, $\Gamma$, and Vega.
\State $T_{\mathrm{yr}} \gets \tau/365$.
\State $P_+^S \gets \mathrm{LR}(S(1+h_S),K,\sigma,T_{\mathrm{yr}},r)$ and $P_-^S \gets \mathrm{LR}(S(1-h_S),K,\sigma,T_{\mathrm{yr}},r)$.
\State $P_0 \gets \mathrm{LR}(S,K,\sigma,T_{\mathrm{yr}},r)$.
\State $P_+^\sigma \gets \mathrm{LR}(S,K,\sigma+h_\sigma,T_{\mathrm{yr}},r)$ and $P_-^\sigma \gets \mathrm{LR}(S,K,\sigma-h_\sigma,T_{\mathrm{yr}},r)$.
\State $\Delta \gets (P_+^S-P_-^S)/(2h_SS)$.
\State $\Gamma \gets (P_+^S-2P_0+P_-^S)/(h_SS)^2$.
\State $\mathrm{Vega} \gets 0.01(P_+^\sigma-P_-^\sigma)/(2h_\sigma)$.
\State \Return $(\Delta,\Gamma,\mathrm{Vega})$.
\end{algorithmic}
\end{algorithm}

\FloatBarrier
\begin{table}[H]
\centering
\caption{Entry prices and short-position outcomes for the LLY put and call over 1{,}000 simulated paths from April 28 to May 29, 2026. The horizon had 31 calendar days and 22 trading-day steps. Option values and P\&L are in dollars per share.}
\label{tab:lly_scenario}
\begin{tabular}{lrr}\toprule
Statistic & Short put & Short call \\\midrule
Initial spot & \multicolumn{2}{c}{873.83} \\
Strike & 825.00 & 940.00 \\
Market delta & -0.303 & +0.309 \\
Market IV & 44.4\% & 44.0\% \\
Market mid & 23.30 & 20.76 \\
Model entry value & 21.95 & 18.96 \\
Model value minus market mid & -1.35 & -1.80 \\
Mean P\&L & 16.39 & 10.61 \\
Monte Carlo SE of mean & 0.63 & 0.91 \\
Median P\&L & 23.30 & 20.76 \\
Standard deviation & 19.86 & 28.66 \\
5\% quantile & -27.91 & -51.96 \\
5\% expected shortfall & -56.84 & -94.18 \\
Premium retained & 80.7\% & 79.3\% \\
Wilson 95\% interval & 78.1--83.0\% & 76.7--81.7\% \\
Delta-heuristic premium retained & 69.7\% & 69.1\% \\
Simulated minus heuristic & +11.0 pp & +10.2 pp \\
\bottomrule\end{tabular}

\end{table}

\FloatBarrier
\begin{figure}[H]
\centering
\includegraphics[width=\textwidth]{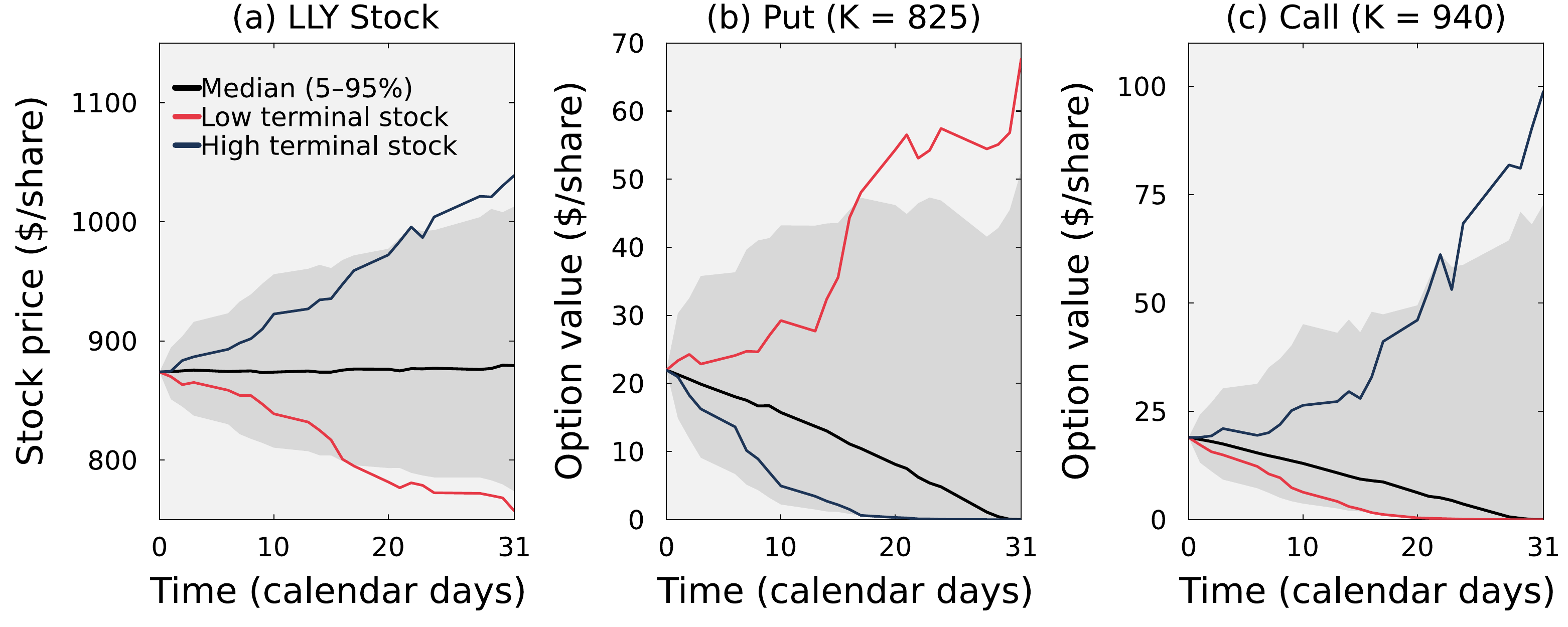}
\caption{Simulated evolution of LLY stock prices (a), put values (b), and call values (c) along 1{,}000 paths between April 28 and May 29, 2026. The fitted surfaces included observations from this period, so these paths illustrate the model and are not an out-of-sample forecast. Time is measured in calendar days from April 28. Black curves show pointwise medians, and shading contains the middle 90\% of values. Red and blue curves show medians of the paths whose terminal stock price fell in the 1st--5th and 95th--99th percentiles. Option values are in dollars per share and give the modeled cost of buying back the option for a short seller. Entry premiums were \$23.30 per share for the put and \$20.76 for the call (Supplementary Table~\ref{tab:lly_scenario}).}
\label{fig:lly_paths}
\end{figure}

\begin{figure}[H]
\centering
\includegraphics[width=\textwidth]{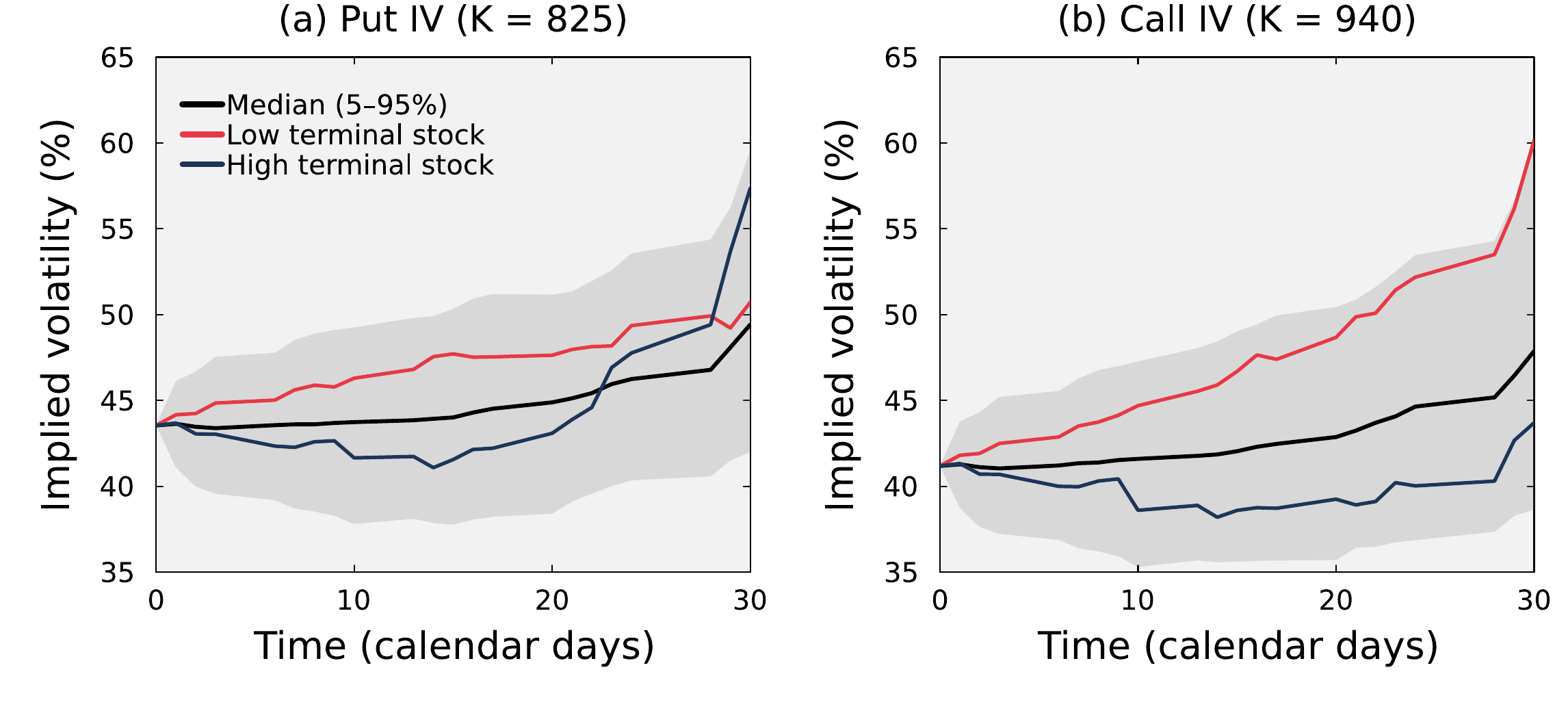}
\caption{Simulated evolution of annualized IV at the LLY put strike (a) and call strike (b). Both panels express IV as percentages on the same vertical scale. Curves show pointwise medians across all 1{,}000 paths and within the terminal-stock groups in Supplementary Fig.~\ref{fig:lly_paths}. Shading shows the 5th--95th percentile band across all paths. Time is measured in calendar days from April 28, 2026. The curves stop before expiration because the payoff at expiration does not depend on IV.}
\label{fig:lly_iv}
\end{figure}

\begin{figure}[H]
\centering
\includegraphics[width=\textwidth]{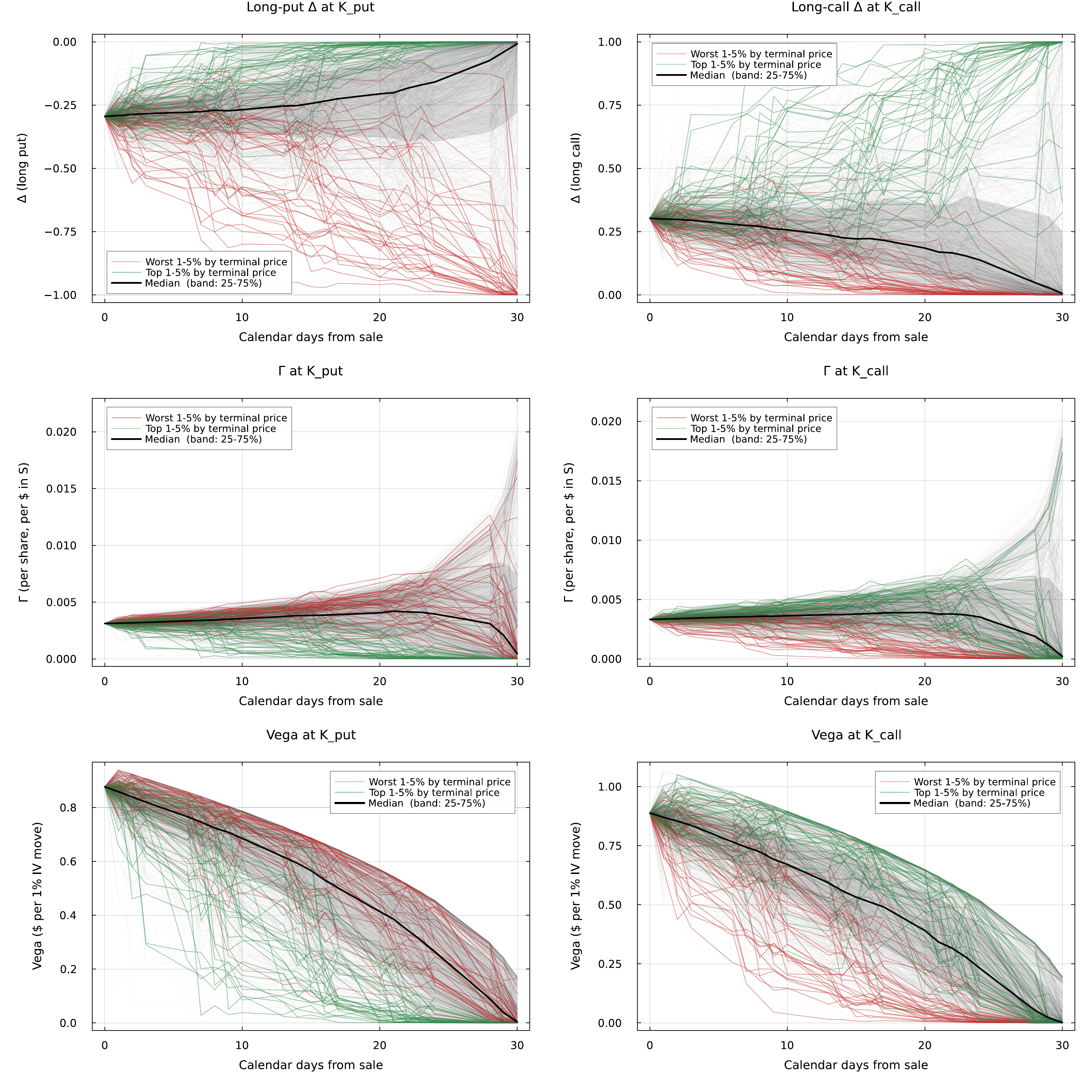}
\caption{Simulated evolution of LLY long-option delta, gamma, and Vega along the stock paths. Sensitivities come from the 201-step Leisen--Reimer tree. Short-position Greeks have the opposite sign.}
\label{fig:lly_greeks}
\end{figure}

\clearpage
\section{IV Ablation and Strike Audit}\label{sec:supp_ablation}

At the fixed primary horizon, ten trading days after entry, we compared hypothetical short-position P\&L under the five IV variants on shared stock paths (Table~\ref{tab:dynamic_ablation}). The paired standard errors in the table apply to mean differences. For the expected-shortfall change from the uncoupled to the full factor, each of 5{,}000 bootstrap replicates resampled paired paths within each seed and pooled them across seeds. The standard deviation of these pooled changes gave the standard error. The three evaluation seeds used the same fitted models and pilot constants. We reported them separately to show whether the differences and the numerical checks held across simulations (Table~\ref{tab:ablation_seeds}).

\begin{table}[H]
\centering
\caption{Hypothetical short-position profit and loss (P\&L) at the fixed ten-trading-day horizon, on May 12, 2026, with 17 calendar DTE remaining. Each row uses 3{,}000 paths. Monetary values are in dollars per share. The columns $\Delta$P\&L (paired P\&L difference) and mean $|\Delta P|$ (mean absolute price difference) compare each variant with the direct surface on identical paths. Parentheses give the paired Monte Carlo standard error of mean $\Delta$P\&L. A more negative 5\% expected shortfall (ES$_{5\%}$) means larger losses in the lower tail. Paired bootstrap standard errors of the change in ES$_{5\%}$ from the uncoupled to the full factor were \$0.15 for the GS put, \$0.24 for the GS call, \$0.13 for the LLY put, and \$0.18 for the LLY call.}\label{tab:dynamic_ablation}
\small
\begin{tabular}{llrrrr}
\toprule
Contract & IV variant & Mean P\&L & ES$_{5\%}$ & $\Delta$P\&L (SE) & Mean $|\Delta P|$ \\
\midrule
GS put & Frozen IV & 0.67 & -63.08 & +0.28 (0.01) & 0.51 \\
GS put & Direct surface & 0.39 & -62.59 & +0.00 (0.00) & 0.00 \\
GS put & Mean reversion & 0.64 & -62.95 & +0.25 (0.01) & 0.42 \\
GS put & Uncoupled factor & 0.70 & -62.98 & +0.31 (0.04) & 1.75 \\
GS put & Full factor & 0.34 & -64.60 & -0.05 (0.04) & 1.77 \\
\midrule
GS call & Frozen IV & -0.03 & -71.07 & +0.87 (0.01) & 0.87 \\
GS call & Direct surface & -0.89 & -72.76 & +0.00 (0.00) & 0.00 \\
GS call & Mean reversion & -0.21 & -71.42 & +0.69 (0.01) & 0.69 \\
GS call & Uncoupled factor & -0.12 & -71.53 & +0.77 (0.04) & 1.93 \\
GS call & Full factor & 0.19 & -69.82 & +1.09 (0.04) & 1.96 \\
\midrule
LLY put & Frozen IV & 7.15 & -35.54 & +1.33 (0.01) & 1.33 \\
LLY put & Direct surface & 5.83 & -36.17 & +0.00 (0.00) & 0.00 \\
LLY put & Mean reversion & 6.96 & -35.53 & +1.13 (0.01) & 1.13 \\
LLY put & Uncoupled factor & 7.01 & -35.70 & +1.18 (0.04) & 2.04 \\
LLY put & Full factor & 6.66 & -38.27 & +0.83 (0.04) & 1.96 \\
\midrule
LLY call & Frozen IV & 6.31 & -47.30 & +2.23 (0.01) & 2.23 \\
LLY call & Direct surface & 4.08 & -50.42 & +0.00 (0.00) & 0.00 \\
LLY call & Mean reversion & 5.90 & -47.95 & +1.82 (0.00) & 1.82 \\
LLY call & Uncoupled factor & 5.95 & -48.48 & +1.87 (0.04) & 2.36 \\
LLY call & Full factor & 6.26 & -45.18 & +2.18 (0.04) & 2.41 \\
\midrule
\end{tabular}

\end{table}
\FloatBarrier

\begin{table}[H]
\centering
\caption{Comparison of the full factor with the direct surface after ten trading days, with numerical diagnostics, for each evaluation seed. Each row uses 1{,}000 paths. Monetary values are in dollars per share. Mean $|\Delta P|$, mean $\Delta$P\&L, and its paired standard error compare the full factor with the direct surface. Outside is the percentage of path dates before expiration on which strike/spot lay outside the 0.8 to 1.2 range of the calibration data. Floor is the largest percentage of path dates after entry at the variance floor, taken across the five IV variants. Tree change is the largest absolute price change from 201 to 401 tree steps across the five variants. It uses the first twenty paths after ten and twenty trading days.}\label{tab:ablation_seeds}
\resizebox{\textwidth}{!}{\begin{tabular}{llrrrrrr}
\toprule
Contract & Seed & Mean $|\Delta P|$ & Mean $\Delta$P\&L & Paired SE & Outside (\%) & Floor (\%) & Tree change (\$) \\
\midrule
GS put & 20260429 & 1.82 & -0.08 & 0.07 & 0.7 & 0.0 & 0.0013 \\
GS put & 20260430 & 1.76 & -0.00 & 0.07 & 0.7 & 0.0 & 0.0013 \\
GS put & 20260431 & 1.73 & -0.06 & 0.07 & 0.8 & 0.0 & 0.0017 \\
GS call & 20260429 & 1.97 & +1.07 & 0.08 & 1.9 & 0.0 & 0.0000 \\
GS call & 20260430 & 1.99 & +1.17 & 0.08 & 2.1 & 0.0 & 0.0000 \\
GS call & 20260431 & 1.92 & +1.03 & 0.08 & 1.9 & 0.0 & 0.0000 \\
LLY put & 20260429 & 2.01 & +0.83 & 0.07 & 0.8 & 0.0 & 0.0014 \\
LLY put & 20260430 & 1.96 & +0.85 & 0.07 & 0.8 & 0.0 & 0.0014 \\
LLY put & 20260431 & 1.90 & +0.81 & 0.07 & 0.7 & 0.0 & 0.0013 \\
LLY call & 20260429 & 2.44 & +2.21 & 0.07 & 2.5 & 0.0 & 0.0000 \\
LLY call & 20260430 & 2.43 & +2.22 & 0.07 & 2.9 & 0.0 & 0.0000 \\
LLY call & 20260431 & 2.37 & +2.12 & 0.07 & 2.6 & 0.0 & 0.0000 \\
\bottomrule
\end{tabular}
}
\end{table}

Because the tree priced each option separately, nothing forced the prices to be consistent across strikes. The strike audit therefore checked necessary conditions for that consistency at 401 tree steps (Table~\ref{tab:strike_audit}). The grid had eleven fixed strikes at the scenario expiration, evaluated at entry and on fifty stock paths after ten and twenty trading days. As the stock moved, some strikes left the moneyness range of the calibration data, so the audit included extrapolated surface values. Each variant and ticker supplied 2{,}222 bound checks, 2{,}020 checks per adjacent-strike condition, and 1{,}818 convexity checks across calls and puts. Violation counts were identical at 201 and 401 steps (Table~\ref{tab:strike_audit_depth}). The saved output lists each test location and violation size.

\begin{table}[H]
\centering
\caption{Strike-condition violations larger than \$0.01 per share at 401 tree steps, by ticker and IV variant. Each entry gives violations/tests. The vertical check tests the upper bound on the price difference between options at adjacent strikes. Max reports the largest positive deviation across the four checks, including deviations below the reporting tolerance.}\label{tab:strike_audit}
\resizebox{\textwidth}{!}{\begin{tabular}{llrrrrr}
\toprule
Ticker & Variant & Bounds & Monotonicity & Vertical & Convexity & Max (\$) \\
\midrule
GS & Frozen IV & 0/2222 & 0/2020 & 0/2020 & 0/1818 & 0.008 \\
GS & Direct surface & 0/2222 & 68/2020 & 68/2020 & 178/1818 & 0.113 \\
GS & Mean reversion & 0/2222 & 2/2020 & 0/2020 & 0/1818 & 0.014 \\
GS & Uncoupled factor & 0/2222 & 3/2020 & 0/2020 & 0/1818 & 0.014 \\
GS & Full factor & 0/2222 & 3/2020 & 0/2020 & 0/1818 & 0.013 \\
LLY & Frozen IV & 0/2222 & 0/2020 & 0/2020 & 0/1818 & 0.000 \\
LLY & Direct surface & 0/2222 & 0/2020 & 0/2020 & 0/1818 & 0.000 \\
LLY & Mean reversion & 0/2222 & 0/2020 & 0/2020 & 0/1818 & 0.000 \\
LLY & Uncoupled factor & 0/2222 & 0/2020 & 0/2020 & 0/1818 & 0.000 \\
LLY & Full factor & 0/2222 & 0/2020 & 0/2020 & 0/1818 & 0.000 \\
\bottomrule
\end{tabular}
}
\end{table}

\begin{table}[H]
\centering
\caption{Strike-condition checks at 201 and 401 tree steps, by ticker, IV variant, and check. Viol.\ 201 and Viol.\ 401 count violations larger than \$0.01 per share at each depth. Tests is the number of checks per variant, which was the same at both depths. Max 201 and Max 401 give the largest positive deviation for that check, in dollars per share, including deviations below the reporting tolerance.}\label{tab:strike_audit_depth}
\small
\setlength{\tabcolsep}{5pt}
\begin{tabular}{lllrrrrr}
\toprule
Ticker & Variant & Check & Viol.\ 201 & Viol.\ 401 & Tests & Max 201 (\$) & Max 401 (\$) \\
\midrule
GS & Frozen IV & Bounds & 0 & 0 & 2222 & 0.000 & 0.000 \\
GS & Frozen IV & Monotonicity & 0 & 0 & 2020 & 0.008 & 0.008 \\
GS & Frozen IV & Vertical & 0 & 0 & 2020 & 0.000 & 0.000 \\
GS & Frozen IV & Convexity & 0 & 0 & 1818 & 0.000 & 0.000 \\
GS & Direct surface & Bounds & 0 & 0 & 2222 & 0.000 & 0.000 \\
GS & Direct surface & Monotonicity & 68 & 68 & 2020 & 0.077 & 0.077 \\
GS & Direct surface & Vertical & 68 & 68 & 2020 & 0.074 & 0.073 \\
GS & Direct surface & Convexity & 178 & 178 & 1818 & 0.113 & 0.113 \\
GS & Mean reversion & Bounds & 0 & 0 & 2222 & 0.000 & 0.000 \\
GS & Mean reversion & Monotonicity & 2 & 2 & 2020 & 0.014 & 0.014 \\
GS & Mean reversion & Vertical & 0 & 0 & 2020 & 0.000 & 0.000 \\
GS & Mean reversion & Convexity & 0 & 0 & 1818 & 0.000 & 0.000 \\
GS & Uncoupled factor & Bounds & 0 & 0 & 2222 & 0.000 & 0.000 \\
GS & Uncoupled factor & Monotonicity & 3 & 3 & 2020 & 0.014 & 0.014 \\
GS & Uncoupled factor & Vertical & 0 & 0 & 2020 & 0.000 & 0.000 \\
GS & Uncoupled factor & Convexity & 0 & 0 & 1818 & 0.000 & 0.000 \\
GS & Full factor & Bounds & 0 & 0 & 2222 & 0.000 & 0.000 \\
GS & Full factor & Monotonicity & 3 & 3 & 2020 & 0.013 & 0.013 \\
GS & Full factor & Vertical & 0 & 0 & 2020 & 0.000 & 0.000 \\
GS & Full factor & Convexity & 0 & 0 & 1818 & 0.000 & 0.000 \\
LLY & Frozen IV & Bounds & 0 & 0 & 2222 & 0.000 & 0.000 \\
LLY & Frozen IV & Monotonicity & 0 & 0 & 2020 & 0.000 & 0.000 \\
LLY & Frozen IV & Vertical & 0 & 0 & 2020 & 0.000 & 0.000 \\
LLY & Frozen IV & Convexity & 0 & 0 & 1818 & 0.000 & 0.000 \\
LLY & Direct surface & Bounds & 0 & 0 & 2222 & 0.000 & 0.000 \\
LLY & Direct surface & Monotonicity & 0 & 0 & 2020 & 0.000 & 0.000 \\
LLY & Direct surface & Vertical & 0 & 0 & 2020 & 0.000 & 0.000 \\
LLY & Direct surface & Convexity & 0 & 0 & 1818 & 0.000 & 0.000 \\
LLY & Mean reversion & Bounds & 0 & 0 & 2222 & 0.000 & 0.000 \\
LLY & Mean reversion & Monotonicity & 0 & 0 & 2020 & 0.000 & 0.000 \\
LLY & Mean reversion & Vertical & 0 & 0 & 2020 & 0.000 & 0.000 \\
LLY & Mean reversion & Convexity & 0 & 0 & 1818 & 0.000 & 0.000 \\
LLY & Uncoupled factor & Bounds & 0 & 0 & 2222 & 0.000 & 0.000 \\
LLY & Uncoupled factor & Monotonicity & 0 & 0 & 2020 & 0.000 & 0.000 \\
LLY & Uncoupled factor & Vertical & 0 & 0 & 2020 & 0.000 & 0.000 \\
LLY & Uncoupled factor & Convexity & 0 & 0 & 1818 & 0.000 & 0.000 \\
LLY & Full factor & Bounds & 0 & 0 & 2222 & 0.000 & 0.000 \\
LLY & Full factor & Monotonicity & 0 & 0 & 2020 & 0.000 & 0.000 \\
LLY & Full factor & Vertical & 0 & 0 & 2020 & 0.000 & 0.000 \\
LLY & Full factor & Convexity & 0 & 0 & 1818 & 0.000 & 0.000 \\
\bottomrule
\end{tabular}

\end{table}
\FloatBarrier

\section{Details of the Tests on Later Dates}\label{sec:chronological_supp}

The surface tests used different data and fitting rules. The first refitted the two-input sector networks from scratch on the first $k$ of 58 snapshots and tested the next, for $k=5,\ldots,57$, using the calibration architecture and training schedule. Each fold standardized inputs from its training rows alone. An independent reconstruction reproduced the expanding-window dataset's 986{,}492 filtered observations and confirmed that all passed the July-cutoff fit's two-sided-quote checks. For the July-cutoff fit, we accepted files retrieved between 20:00 UTC on the stored close's session date and 13:30 UTC on the next trading day, the New York close-to-open interval during April--September. We retained the latest eligible file per ticker and session, removed duplicate option rows, and excluded twenty intraday files. Quotes required finite positive bids, asks no lower than bids, and positive displayed sizes. Surface fitting retained the calibration IV and moneyness filters and required positive DTE recomputed from expiration and session date. Training also excluded retrievals after July 31, 2026, UTC. The 68 training sessions ran from April 13 through July 31 and supplied 1{,}173{,}538 observations. The six networks used seed 42 and the calibration training schedule of at most 2{,}000 epochs. We retained their final weights and ticker levels and used training-only standardization for both fitting and testing. Testing used 425{,}061 observations from 23 sessions between August 4 and September 4; August 3 and 20 had no snapshots. Tables~\ref{tab:walk_forward_extended} and~\ref{tab:chronological_surface} report the fold and July-cutoff errors separately, and Table~\ref{tab:chronological_data_audit} records the data checks.

\begin{table}[H]
\centering
\caption{Next-date IV errors of the sector surfaces across the 53 expanding-window folds, by test month. Each fold fitted earlier capture dates and evaluated the next captured date using its observed stock price. The RMSE is in volatility percentage points. Median, minimum, and maximum are taken across the folds in each month. The last row summarizes the RMSEs of all 53 folds.}\label{tab:walk_forward_extended}
\begin{tabular}{lrrrr}
\toprule
Test month & Folds & Median RMSE & Minimum & Maximum \\
\midrule
April & 6 & 12.72 & 7.79 & 17.45 \\
May & 17 & 9.62 & 8.65 & 11.30 \\
June & 18 & 10.02 & 8.55 & 19.25 \\
July & 12 & 9.76 & 7.68 & 10.51 \\
\midrule
All folds & 53 & 9.92 & 7.68 & 19.25 \\
\bottomrule\end{tabular}

\end{table}

\begin{table}[H]
\centering
\caption{Training and held-out IV errors of the July-cutoff sector surfaces, fitted through July 31. Training used 68 underlying sessions. Testing used 23 later sessions from August 4 through September 4. Each RMSE pools observations within the indicated group. Bias is predicted minus observed IV, in percentage points. The last two rows show GS and LLY alone.}\label{tab:chronological_surface}
\resizebox{\textwidth}{!}{\begin{tabular}{lrrrrr}
\toprule
Group & Train rows & Test rows & Train RMSE & Test RMSE & Test bias \\
\midrule
ETF & 503,487 & 186,987 & 7.03 & 6.59 & +3.08 \\
Financials & 83,005 & 27,853 & 7.40 & 7.81 & +2.50 \\
Energy & 34,817 & 10,956 & 8.15 & 9.40 & +2.45 \\
Retail & 50,776 & 15,096 & 10.02 & 11.77 & +1.83 \\
Healthcare & 126,818 & 45,472 & 12.14 & 12.21 & +3.77 \\
Tech & 374,635 & 138,697 & 14.70 & 22.91 & +14.81 \\
Pooled & 1,173,538 & 425,061 & 10.79 & 14.75 & +6.88 \\
\midrule
GS & 41,119 & 14,330 & 6.78 & 6.30 & +0.88 \\
LLY & 39,242 & 13,753 & 9.47 & 9.96 & +4.61 \\
\bottomrule\end{tabular}
}
\end{table}

\begin{table}[H]
\centering
\caption{Data checks for the expanding-window and July-cutoff experiments. The expanding-window rows come from an independent reconstruction of the saved row counts. The July-cutoff rows come from the capture audit. Every excluded file had a retained file for the same ticker and session.}\label{tab:chronological_data_audit}
\begin{tabular}{llr}
\toprule
Experiment & Check & Value \\
\midrule
Expanding-window folds & Capture dates & 58 \\
 & Option files & 1,790 \\
 & Accepted observations & 986,492 \\
 & Crossed or missing asks & 0 \\
 & Missing or nonpositive bid or ask sizes & 0 \\
\midrule
July-cutoff surfaces & Downloaded files & 2,833 \\
 & Files retrieved during trading hours (excluded) & 20 \\
 & Superseded files (excluded) & 0 \\
 & Files retained & 2,813 \\
 & Training observations (68 sessions) & 1,173,538 \\
 & Test observations (23 sessions) & 425,061 \\
\bottomrule\end{tabular}

\end{table}

The forecast protocol fixed contract selection, simulation, and scoring before aggregate errors were inspected; its maturity amendment responded to missing outcomes. Within the monthly and shorter-maturity ranges, we chose expirations nearest 31 and 14 calendar DTE, respectively, then strikes nearest $K/S=0.95$ for puts and $1.05$ for calls. Ties favored earlier expirations and lower strikes. Missing exact-contract endpoints remained missing, without replacing contracts or dates. Monthly contracts moved between the collector's 14- and 30-day targets after five sessions and were no longer saved; shorter contracts moved from the 14-day target toward the 7-day target and usually remained available (Table~\ref{tab:forecast_availability}). The observed-stock diagnostic required every intervening close and therefore used a smaller subset shared by all variants. Simulations used fixed ticker-and-origin seeds and the 2014--2024 JumpHMM fit, starting from a stationary state with no jump episode in progress. A separate 10{,}000-path pilot per ticker, reused at every origin, fixed the shift to the assumed 10\% annualized log-growth target and the stock-return mean and standard deviation. The random walk used zero log drift and the pilot return standard deviation. The observed-stock diagnostic standardized realized returns with the same pilot constants, and endpoint repricing used the 201-step LR tree. We selected the earliest GS shorter-cohort origin with both five-session outcomes for the example figure, independently of forecast errors. For endpoint values $P_1,\ldots,P_N$ on $N=N_p$ paths and observed midpoint $y$, the empirical continuous ranked probability score~\cite{gneiting2007} was given by:
\begin{equation}\label{eq:forecast_crps}
\mathrm{CRPS}=\frac{1}{N}\sum_{j=1}^{N}|P_j-y|
-\frac{1}{2N^2}\sum_{j=1}^{N}\sum_{l=1}^{N}|P_j-P_l|,
\end{equation}
Smaller CRPS indicated better forecasts. Coverage was the fraction of observed midpoints inside the central 90\% interval of simulated values; interval width measured its spread. For each metric, we averaged its matched-contract contributions within each ticker, cohort, horizon, and origin, then averaged origins. For RMSE, we took the square root of the resulting mean squared error. This gave dates equal weight.

\begin{table}[H]
\centering
\caption{Availability of exact-contract outcomes by cohort, ticker, and horizon. Horizons are in trading sessions. Selected counts all origin-selected contracts whose endpoint fell within August 4--September 4, 2026, including endpoints on missing capture dates. Matched counts the selected contracts with an exact-symbol quote at the endpoint. Full stock path counts selected contracts with all intervening stock observations, whether or not the endpoint option quote was available.}\label{tab:forecast_availability}
\resizebox{\textwidth}{!}{\begin{tabular}{llrrrrrr}
\toprule
Cohort & Ticker & Horizon & Origins & Selected & Matched & Missing & Full stock path \\
\midrule
monthly & GS & 1 & 22 & 44 & 34 & 10 & 42 \\
monthly & GS & 5 & 18 & 36 & 0 & 36 & 26 \\
monthly & LLY & 1 & 22 & 44 & 34 & 10 & 42 \\
monthly & LLY & 5 & 18 & 36 & 0 & 36 & 26 \\
short & GS & 1 & 22 & 44 & 42 & 2 & 42 \\
short & GS & 5 & 18 & 36 & 34 & 2 & 26 \\
short & LLY & 1 & 22 & 44 & 41 & 3 & 42 \\
short & LLY & 5 & 18 & 36 & 30 & 6 & 26 \\
\bottomrule\end{tabular}
}
\end{table}

\begin{table}[H]
\centering
\caption{One-session option forecasts for the monthly cohort under the five IV variants. Scores first average matched contracts within each ticker and origin and then average dates. Monetary quantities (MAE, CRPS, and central 90\% interval width) are in dollars per share. Coverage is the fraction of observed midpoints inside that interval, computed within each origin and then averaged across dates.}\label{tab:forecast_one_session}
\resizebox{\textwidth}{!}{\begin{tabular}{llrrrrrr}
\toprule
Ticker & IV variant & Dates & Quotes & MAE & CRPS & Coverage (\%) & Width \\
\midrule
GS & Frozen IV & 17 & 34 & 3.80 & 2.65 & 85.3 & 15.15 \\
GS & Direct surface & 17 & 34 & 3.79 & 2.66 & 82.4 & 14.68 \\
GS & Mean reversion & 17 & 34 & 3.80 & 2.65 & 85.3 & 15.15 \\
GS & Uncoupled factor & 17 & 34 & 3.80 & 2.65 & 88.2 & 16.05 \\
GS & Full factor & 17 & 34 & 3.79 & 2.65 & 94.1 & 15.64 \\
LLY & Frozen IV & 17 & 34 & 6.11 & 4.52 & 76.5 & 17.32 \\
LLY & Direct surface & 17 & 34 & 6.16 & 4.52 & 76.5 & 17.55 \\
LLY & Mean reversion & 17 & 34 & 6.11 & 4.52 & 76.5 & 17.32 \\
LLY & Uncoupled factor & 17 & 34 & 6.11 & 4.48 & 73.5 & 18.24 \\
LLY & Full factor & 17 & 34 & 6.08 & 4.51 & 73.5 & 17.71 \\
\bottomrule\end{tabular}
}
\end{table}

\begin{table}[H]
\centering
\caption{Five-session option errors in the shorter-maturity cohort with IV updated along the observed stock path. Because this diagnostic uses the realized stock path, it is not a forecast. All variants use the same complete-path subset. Monetary quantities and aggregation follow Table~\ref{tab:forecast_main}. Simulated-path MAE scores the same origins and quotes with the simulated stock paths of the joint forecast. Comparing the MAE columns shows the effect of replacing simulated stock paths with observed paths.}\label{tab:forecast_conditional}
\resizebox{\textwidth}{!}{\begin{tabular}{llrrrrrrr}
\toprule
Ticker & IV variant & Dates & Quotes & MAE & CRPS & Coverage (\%) & Width & Simulated-path MAE \\
\midrule
GS & Frozen IV & 13 & 26 & 1.07 & 1.07 & 0.0 & 0.00 & 4.67 \\
GS & Direct surface & 13 & 26 & 1.45 & 1.45 & 0.0 & 0.00 & 5.32 \\
GS & Mean reversion & 13 & 26 & 1.08 & 1.08 & 0.0 & 0.00 & 4.72 \\
GS & Uncoupled factor & 13 & 26 & 1.07 & 0.71 & 92.3 & 3.37 & 4.70 \\
GS & Full factor & 13 & 26 & 1.23 & 0.90 & 50.0 & 2.72 & 4.62 \\
LLY & Frozen IV & 13 & 24 & 1.74 & 1.74 & 0.0 & 0.00 & 10.66 \\
LLY & Direct surface & 13 & 24 & 2.25 & 2.25 & 0.0 & 0.00 & 10.72 \\
LLY & Mean reversion & 13 & 24 & 1.74 & 1.74 & 0.0 & 0.00 & 10.67 \\
LLY & Uncoupled factor & 13 & 24 & 1.73 & 1.35 & 84.6 & 4.40 & 10.66 \\
LLY & Full factor & 13 & 24 & 1.95 & 1.53 & 65.4 & 3.54 & 10.61 \\
\bottomrule\end{tabular}
}
\end{table}

\begin{table}[H]
\centering
\caption{Scores of the original 1{,}000-path stock forecasts from JumpHMM and a random walk at the observed endpoint dates. The MAE scores forecast means. The random walk has zero log drift and the fixed pilot's one-session return standard deviation. Monetary errors are in dollars per share. Coverage refers to central 90\% forecast intervals.}\label{tab:forecast_stock}
\resizebox{\textwidth}{!}{\begin{tabular}{lrllrrr}
\toprule
Ticker & Horizon & Stock model & Dates & MAE & CRPS & Coverage (\%) \\
\midrule
GS & 1 & JumpHMM & 21 & 10.91 & 8.06 & 90.5 \\
GS & 1 & Random walk & 21 & 10.69 & 8.25 & 95.2 \\
GS & 5 & JumpHMM & 17 & 16.85 & 13.05 & 100.0 \\
GS & 5 & Random walk & 17 & 16.17 & 13.79 & 100.0 \\
LLY & 1 & JumpHMM & 21 & 19.26 & 14.80 & 71.4 \\
LLY & 1 & Random walk & 21 & 18.87 & 14.40 & 76.2 \\
LLY & 5 & JumpHMM & 17 & 49.83 & 35.71 & 70.6 \\
LLY & 5 & Random walk & 17 & 48.64 & 34.33 & 70.6 \\
\bottomrule\end{tabular}
}
\end{table}

\begin{table}[H]
\centering
\caption{Errors of American-tree prices computed with the IV reported at the endpoint, for matched options with valid reported IV. This diagnostic uses information from the endpoint, so it is not a forecast. The MAE is in dollars per share. Inside-spread frequencies compare each price with Alpaca's indicative bid and ask.}\label{tab:forecast_repricing}
\begin{tabular}{llrrrrr}
\toprule
Cohort & Ticker & Horizon & Dates & Quotes & MAE & Inside spread (\%) \\
\midrule
monthly & GS & 1 & 17 & 34 & 0.24 & 97.1 \\
monthly & LLY & 1 & 17 & 34 & 0.46 & 100.0 \\
short & GS & 1 & 21 & 42 & 0.16 & 100.0 \\
short & GS & 5 & 17 & 34 & 0.14 & 100.0 \\
short & LLY & 1 & 21 & 41 & 0.32 & 100.0 \\
short & LLY & 5 & 17 & 30 & 0.34 & 100.0 \\
\bottomrule\end{tabular}

\end{table}

\begin{table}[H]
\centering
\caption{Five-session option errors under mean and median point forecasts for the shorter-maturity cohort. Mean MAE and Mean RMSE score the simulated mean, and Median MAE scores the median. The prespecified comparison scores the mean, and the median is an added summary. All errors are in dollars per share and use the same date-weighted options as Table~\ref{tab:forecast_main}.}\label{tab:option_point_summary}
\begin{tabular}{llrrr}
\toprule
Stock & IV variant & Mean MAE & Median MAE & Mean RMSE \\
\midrule
GS & Frozen IV & 4.21 & 2.48 & 5.06 \\
GS & Direct surface & 4.86 & 2.62 & 5.62 \\
GS & Mean reversion & 4.26 & 2.50 & 5.10 \\
GS & Uncoupled factor & 4.24 & 2.48 & 5.08 \\
GS & Full factor & 4.19 & 2.48 & 4.99 \\
LLY & Frozen IV & 11.62 & 11.78 & 15.00 \\
LLY & Direct surface & 11.62 & 11.59 & 14.74 \\
LLY & Mean reversion & 11.62 & 11.77 & 14.98 \\
LLY & Uncoupled factor & 11.61 & 11.80 & 14.97 \\
LLY & Full factor & 11.59 & 11.79 & 14.92 \\
\bottomrule
\end{tabular}

\end{table}

Before interpreting small differences between IV variants, we checked that the forecast prices were reproducible and numerically accurate. Recomputing prices from the saved stock and variance paths reproduced the forecast means, quantiles, and CRPS. Increasing the tree from 201 to 401 steps changed sampled prices by at most \$0.0026 per share (Table~\ref{tab:forecast_numerical_checks}). At the first origin for each ticker, the paired Monte Carlo standard errors of the mean price difference between the full factor and frozen IV ranged from \$0.041 to \$0.061 across the four shorter-maturity options at five sessions (Table~\ref{tab:forecast_numerical_checks}). These standard errors measure simulation precision for a single price difference, not variation across market dates. These checks do not establish a forecasting advantage from the small average MAE reductions (Table~\ref{tab:forecast_main}). In the observed-stock diagnostic, frozen IV, the direct surface, and mean reversion each gave a single price once the stock path was fixed, so their central intervals had zero width. The stochastic factors' intervals reflected IV uncertainty alone, so their coverage measures something different from the coverage of the joint forecasts (Table~\ref{tab:forecast_conditional}). The saved score files include both cohorts, every horizon with outcomes, all five IV variants, and the date-level errors.

\begin{table}[H]
\centering
\caption{Numerical checks of the forecast prices at the first origin, August 4, 2026. Sampled prices repeat the tree calculation with 201 and 401 steps for both cohorts, all five IV variants, calls and puts, and the first ten paths. Price changes and standard errors are in dollars per share. Paired Monte Carlo SE is the standard error of the mean price difference between the full factor and frozen IV across 1{,}000 paths. It applies to the shorter-maturity options at the five-session endpoint.}\label{tab:forecast_numerical_checks}
\begin{tabular}{lrr}
\toprule
Check & GS & LLY \\
\midrule
Sampled prices, 201 and 401 steps & 300 & 300 \\
Largest price change, 201 to 401 steps & 0.0022 & 0.0026 \\
Paired Monte Carlo SE, five-session put & 0.041 & 0.052 \\
Paired Monte Carlo SE, five-session call & 0.051 & 0.061 \\
\bottomrule\end{tabular}

\end{table}

Four diagnostics help interpret the forecasts: IV-factor persistence, median stock-price change, stock volatility, and surface bias (Table~\ref{tab:mechanism_checks}). With $\kappa=15$ and $\Delta t=1/252$, the Euler update in~\eqref{eq:euler} retained a fraction $1-\kappa\Delta t\approx0.94$ of a factor's deterministic gap to a fixed target at each session. This retention gave a half-life of 11.3 sessions, so the update closed 26.4\% of the gap after five sessions and 45.9\% after ten. The five-session rows used the seventeen scored origins per ticker and the matched contracts in Table~\ref{tab:forecast_main}. Their IV changes compared simulated IV five sessions after origin with the observed origin IV, averaging over paths and then contracts. The one-session volatility rows used the stock-benchmark origins in Table~\ref{tab:stock_validation}. Realized values were sample standard deviations of the observed one-session log returns, and EWMA values averaged the adaptive-volatility model's origin estimates. The GS EWMA estimate began from its 2014--2024 sample variance and rose through the April--July 2026 returns before the first August origin. The surface-bias rows break the held-out ticker bias in Table~\ref{tab:chronological_surface} down by underlying session. The report rows average each ticker's daily bias over the five observed sessions before and after its latest report. They exclude the report day, weight each session by its number of observations, and use training and held-out days alike. The public earnings calendar dated LLY's report on August 5, 2026, after the training cutoff, and GS's on July 14, 2026, inside the training window.

\begin{table}[H]
\centering
\caption{Diagnostics of IV-factor persistence, median stock-price change, stock volatility, and surface bias. Five-session rows average the scored shorter-maturity origins and matched options. Changes in IV are in volatility percentage points relative to the observed origin IV. One-session stock rows use the 2025 and August--September 2026 stock-benchmark origins. Surface bias is predicted minus observed IV for the July-cutoff sector surface, in volatility percentage points. Earnings reports fell on August 5, 2026, for LLY and July 14, 2026, for GS. Report rows exclude the report day.}\label{tab:mechanism_checks}
\begin{tabular}{lrr}
\toprule
Quantity & GS & LLY \\
\midrule
\multicolumn{3}{l}{\textit{Five-session forecasts, shorter-maturity cohort}} \\
JumpHMM median price change (\%) & 0.27 & 0.16 \\
JumpHMM log-return standard deviation (\%) & 3.98 & 3.76 \\
Full-factor mean IV minus origin IV (points) & 0.30 & 0.18 \\
Full-factor mean absolute IV change (points) & 2.49 & 2.47 \\
Direct-surface mean absolute IV change (points) & 5.50 & 4.12 \\
\midrule
\multicolumn{3}{l}{\textit{One-session stock return standard deviation (\%)}} \\
JumpHMM pilot & 1.80 & 1.70 \\
Realized, 2025 & 1.97 & 2.57 \\
EWMA at origin, 2025 & 1.94 & 2.52 \\
Realized, August--September 2026 & 1.40 & 2.19 \\
EWMA at origin, August--September 2026 & 2.26 & 2.14 \\
\midrule
\multicolumn{3}{l}{\textit{Held-out July-cutoff surface bias (points)}} \\
August 4 & $-$3.07 & $-$17.90 \\
August 5 & $-$0.96 & $-$0.53 \\
August 6--September 4, mean & 1.20 & 6.17 \\
August 6--September 4, range & $-$0.68 to 3.59 & 2.04 to 8.07 \\
\midrule
\multicolumn{3}{l}{\textit{Surface bias around each ticker's latest report (points)}} \\
Five observed sessions before & $-$3.93 & $-$11.26 \\
Five observed sessions after & $-$0.57 & 5.19 \\
Change & 3.36 & 16.45 \\
\bottomrule
\end{tabular}

\end{table}

The timing of LLY's report, after the training cutoff, suggested an earnings explanation for its held-out bias. We tested this explanation by refitting the July surfaces with the two earnings inputs used in the April holdout (Table~\ref{tab:earnings_refit}). A protocol written before refitting retained the training and test rows, hidden-layer sizes, and training schedule. Both versions used seeds 42, 43, and 44 and underlying-session dates. For each seed, we averaged LLY's signed IV errors over all observations from August 6 through September 4, then took the absolute value. Support required the average of these absolute biases across seeds to fall by at least half, with every earnings-aware seed below every two-input seed. No reduction meant no support; intermediate outcomes meant partial support. The earnings inputs reduced that bias from 6.17 to 3.85 volatility points on average, and every earnings-aware seed fell below every two-input seed. The 38\% reduction fell short of half, so the prespecified verdict was partial. The inputs also reduced LLY's underprediction on August 4 from 17.90 to 12.86 points and held-out RMSE pooled across all tickers from 14.75 to 13.94 points. The held-out RMSE for GS changed by less than 0.1 point. The two-input refits varied by less than 0.05 points across seeds in every reported quantity. Each sector network shared its earnings-response shape, with ticker-specific levels but no ticker-specific earnings parameters.

\begin{table}[H]
\centering
\caption{Bias and RMSE of the July-cutoff sector surfaces refitted with and without earnings inputs. Values are means across training seeds 42, 43, and 44, with seed ranges in parentheses. Bias is mean predicted minus observed IV. The post-report LLY biases were positive for every seed, so their signed and absolute values coincide. Bias and RMSE are in volatility percentage points. The two-input fit with seed 42 is the published surface. The earnings inputs are signed calendar days to the ticker's nearest report, clipped to $[-30,30]$, and absolute days to the nearest same-sector peer report, capped at 30. For ETFs, both inputs equal the same absolute day distance to the nearest report among all non-ETF stocks, capped at 30.}\label{tab:earnings_refit}
\begin{tabular}{lrr}
\toprule
Quantity & Two inputs & With earnings inputs \\
\midrule
LLY bias, August 6--September 4 & 6.17 (6.16 to 6.18) & 3.85 (3.54 to 4.05) \\
LLY bias, August 4 & $-$17.90 ($-$17.92 to $-$17.88) & $-$12.86 ($-$13.04 to $-$12.73) \\
GS bias, August 6--September 4 & 1.21 (1.20 to 1.22) & 0.51 (0.46 to 0.57) \\
LLY held-out RMSE & 9.96 (9.96 to 9.97) & 8.50 (8.36 to 8.59) \\
GS held-out RMSE & 6.31 (6.30 to 6.31) & 6.37 (6.31 to 6.41) \\
Pooled held-out RMSE & 14.75 (14.75 to 14.76) & 13.94 (13.87 to 14.00) \\
Pooled training RMSE & 10.79 (10.78 to 10.79) & 10.35 (10.29 to 10.39) \\
\bottomrule
\end{tabular}

\end{table}
\FloatBarrier

\section{Stock Forecast Evaluation}\label{sec:stock_validation_supp}

Before comparing stock forecast methods, we checked the saved JumpHMM forecasts in two ways. The first check concerned how returns were scaled. The saved GS and LLY models used daily steps of $1/252$ year and annualized excess log returns. Reconstructing the first-origin forecasts with this scaling reproduced the saved stock means at both one and five sessions. The second check concerned the starting hidden state, which the original forecasts drew from the stationary distribution without using recent returns. In this initialization check, we instead inferred the current hidden state from the available daily returns and advanced it before generating the first future return. The inference tracked both the return state and the number of forced tail-state emissions left in a jump episode. We capped Poisson durations where tail probability was at most $10^{-12}$, assigning that tail mass to the maximum duration. A missing day advanced the hidden-state distribution without an observation likelihood. A return accumulated over several days was never treated as a daily observation. We removed the fixed pilot growth shift from observed annualized returns before computing likelihoods, to match the simulated observation model. We compared this conditional start with the original start on the same 2026 endpoints (Table~\ref{tab:stock_initialization}). Both used three 10{,}000-path runs, with seeds separate from the four-method benchmark. The fitted transitions mixed the hidden states within a few sessions, so the conditional start had little effect on five-session central forecasts. Other exploratory checks, including a fixed 20-return volatility adjustment and Gaussian random walks, are kept in the saved experiment outputs.

\begin{table}[H]
\centering
\caption{Stock forecast errors from the original starting state and from a starting state conditioned on past returns, on identical 2026 endpoints. Each ticker has 21 one-session and 17 five-session origins. The MAE scores the forecast median. Errors are in dollars per share and average three 10{,}000-path runs. Filtered denotes initialization from available past returns, including inference of remaining jump duration.}\label{tab:stock_initialization}
\resizebox{\textwidth}{!}{\begin{tabular}{lrrrrr}
\toprule
Stock & Horizon & Original MAE & Filtered MAE & Original CRPS & Filtered CRPS \\
\midrule
GS & 1 & 10.64 & 10.79 & 7.99 & 8.44 \\
GS & 5 & 16.28 & 16.36 & 12.99 & 13.21 \\
LLY & 1 & 19.10 & 19.08 & 14.73 & 14.62 \\
LLY & 5 & 49.58 & 49.40 & 35.69 & 35.35 \\
\bottomrule
\end{tabular}
}
\end{table}

The benchmark comparison kept JumpHMM's fitted parameters and truncated emissions and tested the adaptive-volatility and direction-predicting models against it. We chose the settings of both models on earlier data only. For each annual validation block from 2019 through 2024, regression training responses ended before the validation year and validation endpoints remained within it. The EWMA variance began with the sample variance of the first sixty training daily returns and then updated using only returns observed by each origin. Decay selection minimized $\tfrac12[\log(h a_t)+y_{t,h}^2/(h a_t)]$, where $y_{t,h}$ was the observed cumulative log return over $h$ sessions. We first averaged this loss within each ticker, validation year, and horizon, then gave the resulting groups equal weight. With the selected decay, the direction-predicting model's regression fitted the next daily return divided by origin volatility. Its four features were the stock's last daily return, its last five- and twenty-return sums, and SPY's last five-return sum. Return sums were divided by the square root of the window length and by the same asset's EWMA volatility at the origin, so SPY's feature used SPY's volatility. Each feature column was then divided by its training root mean square. The ridge objective was the mean squared normalized one-session error plus the penalty times the squared coefficient norm, with no intercept. Penalty selection used $(y_{t,h}-h\widehat\mu_t)^2/(h a_t)$, with the same group weighting as decay selection. The zero-drift limit had the lowest validation loss and set all four coefficients to zero for both tickers (Table~\ref{tab:stock_selection}). The two selection losses measure different things, so their magnitudes cannot be compared. Final coefficients and feature scales used only 2014--2024 data and were saved before scoring this benchmark.

\begin{table}[H]
\centering
\caption{Validation losses used to choose the EWMA decay and ridge penalty of the adaptive-volatility and direction-predicting models. Losses weight ticker, year, and horizon groups equally across the 2019--2024 validation blocks. Lower values are better within each selection task. The selected settings were an EWMA decay of $0.97$ and the zero-drift limit.}\label{tab:stock_selection}
\begin{tabular}{llr}
\toprule
Selection & Setting & Validation loss \\
\midrule
EWMA decay & 0.80 & -2.921492 \\
EWMA decay & 0.90 & -3.009741 \\
EWMA decay & 0.94 & -3.044490 \\
EWMA decay & 0.97 & -3.065558 \\
EWMA decay & 0.99 & -3.064379 \\
Ridge penalty & 0.01 & 1.146523 \\
Ridge penalty & 0.1 & 1.143850 \\
Ridge penalty & 1 & 1.138475 \\
Ridge penalty & 10 & 1.136428 \\
Ridge penalty & 100 & 1.136142 \\
Ridge penalty & 1000 & 1.136112 \\
Ridge penalty & Zero drift & 1.136109 \\
\bottomrule
\end{tabular}

\end{table}

The two test periods differed in how current the stock data were at each origin. Historical close series for GS, LLY, and SPY shared the same recorded session grid, with 2{,}767 closes per ticker in 2014--2024 and 250 in 2025. The 2025 test updated lag features and variance using information through each origin while keeping fitted parameters fixed. The 2026 series began in April after a data gap. Its variance restarted from the 2014--2024 sample estimate and updated through the available April--July daily returns before evaluation. No daily return spanned the gap from December 2025 to April 2026. Within 2026, a daily return required closes on both adjacent trading sessions, and a missing observation left the variance unchanged. Feature windows used the last required number of valid daily returns, which could span more sessions when observations were missing. At evaluated 2026 origins, the most recent return used by the features was up to four calendar days old, whereas every 2025 origin had a return ending on the origin date.

Complete five-session and one-session scores cover all four candidates and both periods, including the direction-predicting rows that the main stock comparison omits (Tables~\ref{tab:stock_validation_five} and~\ref{tab:stock_validation_one}). The direction-predicting and adaptive-volatility rows agree exactly because zero drift was selected. Their small median-error differences from the unchanged-price benchmark are Monte Carlo noise, since the population median of their forecast distribution equals the origin price. The three seeds measure simulation variation alone. The saved outputs include seed-level scores, date-level comparisons, input hashes, and settings.

\begin{table}[H]
\centering
\caption{Complete five-session stock forecast scores for the four selected candidates: JumpHMM, the unchanged-price benchmark, and the adaptive-volatility and direction-predicting models. Each ticker has 245 origins in 2025 and 17 in August--September 2026. The MAE scores medians, and the RMSE scores means. Monetary values are in dollars per share. Coverage and width describe central 90\% intervals. Scores average dates within each of three 10{,}000-path runs and then average the runs. The direction-predicting rows equal the adaptive-volatility rows because earlier-data selection chose zero drift.}\label{tab:stock_validation_five}
\resizebox{\textwidth}{!}{\begin{tabular}{lllrrrrr}
\toprule
Year & Stock & Model & Median MAE & Mean RMSE & CRPS & Coverage (\%) & Width \\
\midrule
2025 & GS & JumpHMM & 22.31 & 28.42 & 16.22 & 84.4 & 84.71 \\
2025 & GS & Unchanged & 22.61 & 28.75 & 22.61 & -- & -- \\
2025 & GS & Adaptive volatility & 22.62 & 28.57 & 16.32 & 89.8 & 94.86 \\
2025 & GS & Direction-predicting & 22.62 & 28.57 & 16.32 & 89.8 & 94.86 \\
2025 & LLY & JumpHMM & 39.10 & 50.16 & 29.60 & 69.9 & 100.40 \\
2025 & LLY & Unchanged & 39.37 & 50.36 & 39.37 & -- & -- \\
2025 & LLY & Adaptive volatility & 39.35 & 50.23 & 28.57 & 84.4 & 151.37 \\
2025 & LLY & Direction-predicting & 39.35 & 50.23 & 28.57 & 84.4 & 151.37 \\
2026 & GS & JumpHMM & 16.25 & 21.71 & 13.03 & 100.0 & 129.05 \\
2026 & GS & Unchanged & 15.51 & 20.71 & 15.51 & -- & -- \\
2026 & GS & Adaptive volatility & 15.48 & 21.07 & 15.58 & 100.0 & 176.18 \\
2026 & GS & Direction-predicting & 15.48 & 21.07 & 15.58 & 100.0 & 176.18 \\
2026 & LLY & JumpHMM & 49.44 & 58.63 & 35.60 & 70.6 & 147.26 \\
2026 & LLY & Unchanged & 49.37 & 58.17 & 49.37 & -- & -- \\
2026 & LLY & Adaptive volatility & 49.31 & 58.35 & 33.81 & 94.1 & 191.38 \\
2026 & LLY & Direction-predicting & 49.31 & 58.35 & 33.81 & 94.1 & 191.38 \\
\bottomrule
\end{tabular}
}
\end{table}

\begin{table}[H]
\centering
\caption{Complete one-session stock forecast scores for the same four candidates. Each ticker has 249 origins in 2025 and 21 in August--September 2026. Units, point summaries, interval definitions, and aggregation follow Supplementary Table~\ref{tab:stock_validation_five}.}\label{tab:stock_validation_one}
\resizebox{\textwidth}{!}{\begin{tabular}{lllrrrrr}
\toprule
Year & Stock & Model & Median MAE & Mean RMSE & CRPS & Coverage (\%) & Width \\
\midrule
2025 & GS & JumpHMM & 9.10 & 12.26 & 6.66 & 87.8 & 34.66 \\
2025 & GS & Unchanged & 9.13 & 12.28 & 9.13 & -- & -- \\
2025 & GS & Adaptive volatility & 9.13 & 12.27 & 6.74 & 93.4 & 42.44 \\
2025 & GS & Direction-predicting & 9.13 & 12.27 & 6.74 & 93.4 & 42.44 \\
2025 & LLY & JumpHMM & 14.03 & 20.43 & 10.67 & 77.5 & 40.23 \\
2025 & LLY & Unchanged & 14.07 & 20.44 & 14.07 & -- & -- \\
2025 & LLY & Adaptive volatility & 14.06 & 20.43 & 10.63 & 93.3 & 67.61 \\
2025 & LLY & Direction-predicting & 14.06 & 20.43 & 10.63 & 93.3 & 67.61 \\
2026 & GS & JumpHMM & 10.67 & 14.34 & 8.00 & 90.5 & 52.39 \\
2026 & GS & Unchanged & 10.59 & 14.23 & 10.59 & -- & -- \\
2026 & GS & Adaptive volatility & 10.60 & 14.30 & 8.72 & 100.0 & 77.21 \\
2026 & GS & Direction-predicting & 10.60 & 14.30 & 8.72 & 100.0 & 77.21 \\
2026 & LLY & JumpHMM & 19.11 & 25.74 & 14.72 & 74.6 & 58.01 \\
2026 & LLY & Unchanged & 19.01 & 25.79 & 19.01 & -- & -- \\
2026 & LLY & Adaptive volatility & 18.98 & 25.78 & 14.47 & 76.2 & 83.93 \\
2026 & LLY & Direction-predicting & 18.98 & 25.78 & 14.47 & 76.2 & 83.93 \\
\bottomrule
\end{tabular}
}
\end{table}
\FloatBarrier

\clearpage
\section{Return Cutoffs and Sensitivity}\label{sec:truncated_emissions}

Every JumpHMM price simulation used the truncated emissions in~\eqref{eq:truncated_emission}, with $b=10$ as the primary cutoff and $b=20$ as the wider sensitivity case. We set both cutoffs as stress assumptions before rerunning the affected experiments, without reference to the largest observed return. The truncation also applied to the state-inference likelihoods, and each experiment recomputed its pilot or ensemble normalization under each cutoff.

To show what the cutoffs mean for daily stock moves, we converted them to daily percentage returns using the pilot growth shift from the tests on later dates (Table~\ref{tab:emission_bounds}). These limits are the widest possible daily returns over all fitted states, not forecast intervals. The removed mass is the probability cut from the standardized Student-$t$ distribution, before state locations and scales are applied. Both models used five degrees of freedom in every state, so the removed mass was the same for GS and LLY. The upper limits nearly matched because each model's highest state had a similar fitted location and scale, 15.30 and 6.77 for GS and 15.00 and 6.81 for LLY, in annualized growth-rate units. The lowest states differed more, so the lower limits did not match. Growth shifts differ between experiments, and so do their daily limits.

\begin{table}[H]
\centering
\caption{Daily return limits under the primary ($b=10$) and wider ($b=20$) cutoffs. The limits include the pilot growth shift from the tests on later dates and the $1/252$ year step. Removed mass is the probability excluded from the original Student-$t$ emission, in percent.}\label{tab:emission_bounds}
\begin{tabular}{lrrrr}
\toprule
Stock & $b$ & Lower return (\%) & Upper return (\%) & Removed mass (\%) \\
\midrule
GS & 10 & -26.47 & 38.99 & 0.0171 \\
GS & 20 & -42.73 & 81.83 & 0.0006 \\
LLY & 10 & -21.23 & 38.98 & 0.0171 \\
LLY & 20 & -38.22 & 82.11 & 0.0006 \\
\bottomrule\end{tabular}

\end{table}

To test whether the results depended on the cutoff, we reran the GS and LLY illustrations, the IV ablation, both forecast cohorts, and the stock benchmarks at each cutoff with the same seeds and inputs. Any differences therefore come only from the changed cutoff and the recomputed normalization constants. We compared short-call losses at expiration in the illustrations and after ten trading days in the full-factor ablation (Table~\ref{tab:tail_sensitivity}). We also compared five-session full-factor option errors in the shorter-maturity cohort and JumpHMM stock scores (Table~\ref{tab:tail_forecasts}).

\begin{table}[H]
\centering
\caption{Sensitivity of short-call P\&L to the return cutoff $b$. Values are in dollars per share. The illustration uses 1{,}000 paths and P\&L at expiration. The full-factor ablation uses 3{,}000 paths and hypothetical P\&L at the fixed ten-trading-day horizon. Expected shortfall (ES$_{5\%}$) averages outcomes at or below the 5\% quantile.}\label{tab:tail_sensitivity}
\begin{tabular}{llrrrr}
\toprule
Experiment & Stock & $b$ & Mean P\&L & 5\% quantile & ES$_{5\%}$ \\
\midrule
Ten-session ablation & GS & 10 & 0.194 & -35.194 & -69.817 \\
Ten-session ablation & GS & 20 & 0.191 & -35.203 & -69.828 \\
Ten-session ablation & LLY & 10 & 6.260 & -20.279 & -45.182 \\
Ten-session ablation & LLY & 20 & 6.263 & -20.273 & -45.174 \\
Terminal illustration & GS & 10 & -2.194 & -86.183 & -130.770 \\
Terminal illustration & GS & 20 & -2.213 & -86.243 & -130.833 \\
Terminal illustration & LLY & 10 & 10.605 & -51.956 & -94.176 \\
Terminal illustration & LLY & 20 & 10.595 & -52.006 & -94.228 \\
\bottomrule\end{tabular}

\end{table}

\begin{table}[H]
\centering
\caption{Sensitivity of five-session option and stock forecast scores to the return cutoff $b$. Option MAE scores the full-factor forecast mean in the shorter-maturity cohort. Stock CRPS scores the JumpHMM forecast distribution in each period. Values are in dollars per share and use the same date and seed weighting as the main comparisons.}\label{tab:tail_forecasts}
\begin{tabular}{llrrr}
\toprule
Quantity & Stock & $b=10$ & $b=20$ & Change \\
\midrule
Option MAE, 2026 & GS & 4.193743 & 4.200236 & +0.006492 \\
Option MAE, 2026 & LLY & 11.587739 & 11.587747 & +0.000008 \\
Stock CRPS, 2025 & GS & 16.221054 & 16.221403 & +0.000349 \\
Stock CRPS, 2025 & LLY & 29.604370 & 29.603983 & -0.000386 \\
Stock CRPS, 2026 & GS & 13.025558 & 13.025238 & -0.000320 \\
Stock CRPS, 2026 & LLY & 35.602434 & 35.601074 & -0.001360 \\
\bottomrule\end{tabular}

\end{table}
\FloatBarrier

\end{document}